\documentclass{ar2e}
 \pdfoutput=1
 
\usepackage{graphicx}
\usepackage{amssymb}
\usepackage{amsmath}
\usepackage{graphicx}
\usepackage{rotating}
\usepackage{subfigure}
\usepackage{setspace} 
\usepackage{natbib}  
\bibpunct{(}{)}{,}{a}{}{,}%

\def\beqa{\begin{eqnarray}}
\def\eeqa{\end{eqnarray}}

\def\beq{\begin{equation}}
\def\eeq{\end{equation}}

\def\sky{{\theta}}
\def\r{\mathbf {r}}
\def\u{\mathbf{u}}
\def\Tsys{{T_{\rm sys}}}
\def\k{\mathbf{k}}
\def\kperp{{k_{\perp}}}
\def\kpar{{k_{\parallel}}}
\def\degree{{^{\circ}}}
\def\f{{\kappa}}

\begin{document}

                   


\jname{..}
\jyear{2009}
\jvol{}
\ARinfo{1056-8700/97/0610-00}

\title{Reionization and Cosmology with 21~cm Fluctuations}

\markboth{Cosmic 21~cm Fluctuations}{Cosmic 21~cm Fluctuations}

\author{Miguel F. Morales
\affiliation{Department of Physics, University of Washington, Seattle, Washington, USA} 
J. Stuart B. Wyithe
\affiliation{School of Physics, University of Melbourne, Parkville, Victoria, Australia}}


\begin{abstract}

Measurement of the spatial distribution of neutral hydrogen via the redshifted 21~cm line promises to revolutionize our knowledge of the epoch of reionization and the first galaxies, and may provide a powerful new tool for observational cosmology from redshifts $1<z<4$. In this review we discuss recent advances in our theoretical understanding of the epoch of reionization (EoR), the application of 21~cm tomography to cosmology and measurements of the dark energy equation of state after reionization, and the instrumentation and observational techniques shared by 21~cm EoR and post reionization cosmology machines. We place particular emphasis on the expected signal and observational capabilities of first generation 21~cm fluctuation instruments.

 
 
\end{abstract}

\maketitle

\section{Evolution of HI in the Universe}

Three major stages in the evolution of our universe are written in the phases of hydrogen. After nucleosynthesis the universe was an ionized plasma of hydrogen and helium. As expansion cooled the universe, hydrogen went through a phase transition, rapidly becoming neutral and releasing the Cosmic Microwave Background light at a redshift of $\sim$1089.
High energy photons produced by the first stars and quasars later reionized the hydrogen in the inter galactic medium (IGM), forcing the universe back through a second extended and patchy phase transition referred to as the epoch of reionization (EoR).

The epoch of reionization was an important milestone in the history
of the universe for two reasons. First, reionization identifies the epoch when astrophysical sources
produced of order one photon per baryon, and so became the dominant
influence on the conditions in the intergalactic medium. Turning this 
around, the study of reionization provides an opportunity to
study the properties of the first galaxies and stars. Since the ionization and temperature of the medium play a significant regulatory role in galaxy formation for all subsequent galaxy evolution, reionization must be understood as part of a complete theory of galaxy formation at any cosmic epoch.

The redshift of the ionized to neutral phase transition is well measured by the cosmic microwave background (CMB, \cite{Komatsu:2009p4359}), but the redshift range of the Epoch of Reionization is significantly less certain. The absence of Gunn-Peterson absorption troughs in the spectra of high redshift quasars indicates that the universe was reionized by a redshift of six and has been highly ionized for the subsequent 12 billion years \citep[see the review by ][]{Fan:2006p3905}. At $z>6$ interpretation of the Lyman-$\alpha$ quasar absorption data becomes more uncertain.

Recent advances in radio instrumentation and techniques will soon make it possible to observe the 21~cm emission of the neutral hydrogen itself. For the EoR this would enable detailed studies of structure formation and the formation of the first galaxies. After reionization, the neutral hydrogen would trace the locations of galaxies and through the power spectrum provide a very promising way of measuring the expansion history of space over redshifts of $1<z<4$.

In this review we concentrate on the astrophysics and cosmology which will be enabled by the first generation EoR and post reionization 21~cm fluctuation experiments. This review should be seen as an installment in a series of reviews by different authors \citep{BL01,Fan:2006p3905, Furlanetto:2006p341}. In particular, the excellent and encyclopedic EoR review by \cite{Furlanetto:2006p341} goes into many details we do not have the space to cover here. In this review we will concentrate on three areas: recent advances in our theoretical understanding of what the EoR may look like, the application of 21~cm tomography to cosmology observations after reionization, and the instrumentation and observational requirements shared by both EoR and post-reionization HI machines.
In our numerical examples,
we adopt the currently standard set of cosmological parameters \citep{Komatsu:2009p4359}\footnote{ $\Omega_{\rm m}=0.24$, $\Omega_{\rm b}=0.04$
and $\Omega_Q=0.76$ for the matter, baryon, and dark energy fractional
density respectively, $h=0.73$ for the dimensionless Hubble constant,
and $\sigma_8=0.81$ for the variance of the linear density field
within regions of radius $8h^{-1}$Mpc.}


\subsection{Current understanding of HI} 

\subsubsection{hydrogen during the reionization era}
\label{overview}
The discovery of
distant quasars over the last decade has facilitated detailed Ly$\alpha$
absorption studies of the state of the high redshift IGM at a time when the universe was as little as a billion years
old \citep{F06}. 
Unfortunately, Ly$\alpha$ absorption can only be used to probe volume averaged neutral fractions
that are smaller than $10^{-3}$ owing to the large cross-section of
the Ly$\alpha$ resonance. As a result, studies of the ionization state of the mean IGM using Ly$\alpha$
absorption become inconclusive in the era of interest for
reionization~\citep[e.g.][]{bolton2007,lidz2007b,becker2007}. 

On the other hand \citet{bolton2007b} have
shown that the observed ionization rate at $z\sim6$ implies an
emissivity that is only just sufficient to have reionized the universe
by that time, indicating that the reionization of
the IGM may have been photon starved. 
Similarly, the small escape fractions found for high
redshift galaxies by several studies \citep[e.g.][]{chen2007}, together 
with the star formation
rates implied by the observed high redshift galaxy population suggests
a photon budget that struggles to have been sufficient to reionize the
universe by $z\sim6$ \citep{gnedin2007b}. If true, these
results imply that while the IGM seems to be highly ionized along the
lines-of-sight towards the highest redshift quasars, the reionization epoch should not have occurred at a redshift substantially
higher than $z\sim7-8$.

Thomson scattering of CMB photons provides a complimentary constraint to the Ly$\alpha$ forest \citep[and also to related studies such as those using Ly$\alpha$ emitters, e.g.][]{ota2008}.
The Thomson scattering optical depth due to free electrons in the IGM probes the integrated
ionization along the line-of-sight between the observer and the surface of last scattering, and thus probes the whole history
over which the universe was reionized. 
The optical depth can therefore be used to estimate the redshift at
which reionization was substantially underway, although it provides only a single number to describe the integrated reionization
history and so is degenerate among a wide range of reionization
scenarios. 
Assuming an instantaneous redshift for reionization, measurements from the WMAP satellite yield $z_{\rm reion}=10.9\pm1.4$~\citep{Komatsu:2009p4359}.

Several authors have considered the constraints available on the
reionization history based on a combination of existing observations. For example
\citet{Choudhury2006} and \citet{PLW09} have modeled the reionization history of hydrogen and compared predictions to a range of
observables, using results for specific physical models and generic parameterised models respectively.
The consistent message from modeling of the reionization history is that reionization must have been
completed by $z>6.5$, but could not have been completed at a redshift substantially higher than $z\sim7-8$ given the value of optical depth to electron scattering. An example reionization history showing the evolution of the mean mass weighted neutral fraction (designated $\bar{x}_{\rm HI}$ throughout this review) in a model that is consistent with current constraints is shown in Figure~\ref{fig_WL} for illustration.

These studies illustrate the limited utility of current observational probes of the IGM for studying the EoR in detail.
Ly$\alpha$ absorption studies become ineffective as reionization is
approached due to the high optical depth of the absorption line, and can only be studied along a few sight lines due to the rarity of bright quasars. The CMB optical depth currently
offers our most robust constraints, but it is restricted to being an
integral, sky averaged measure \citep[though the Planck
satellite may be able to trace some details of the reionization
history,][]{holder2003}. 
What is required to further advance our knowledge of reionization and the 
first galaxies is measurement of the ionization state of the IGM as a function of both
time and space. Such an opportunity is provided by tomography of the redshifted 21~cm emission line \citep[e.g.][]{tozzi2000}.

\subsubsection{hydrogen following the completion of reionization}

\begin{figure}
\begin{center}
\includegraphics[width = 5.25 in]{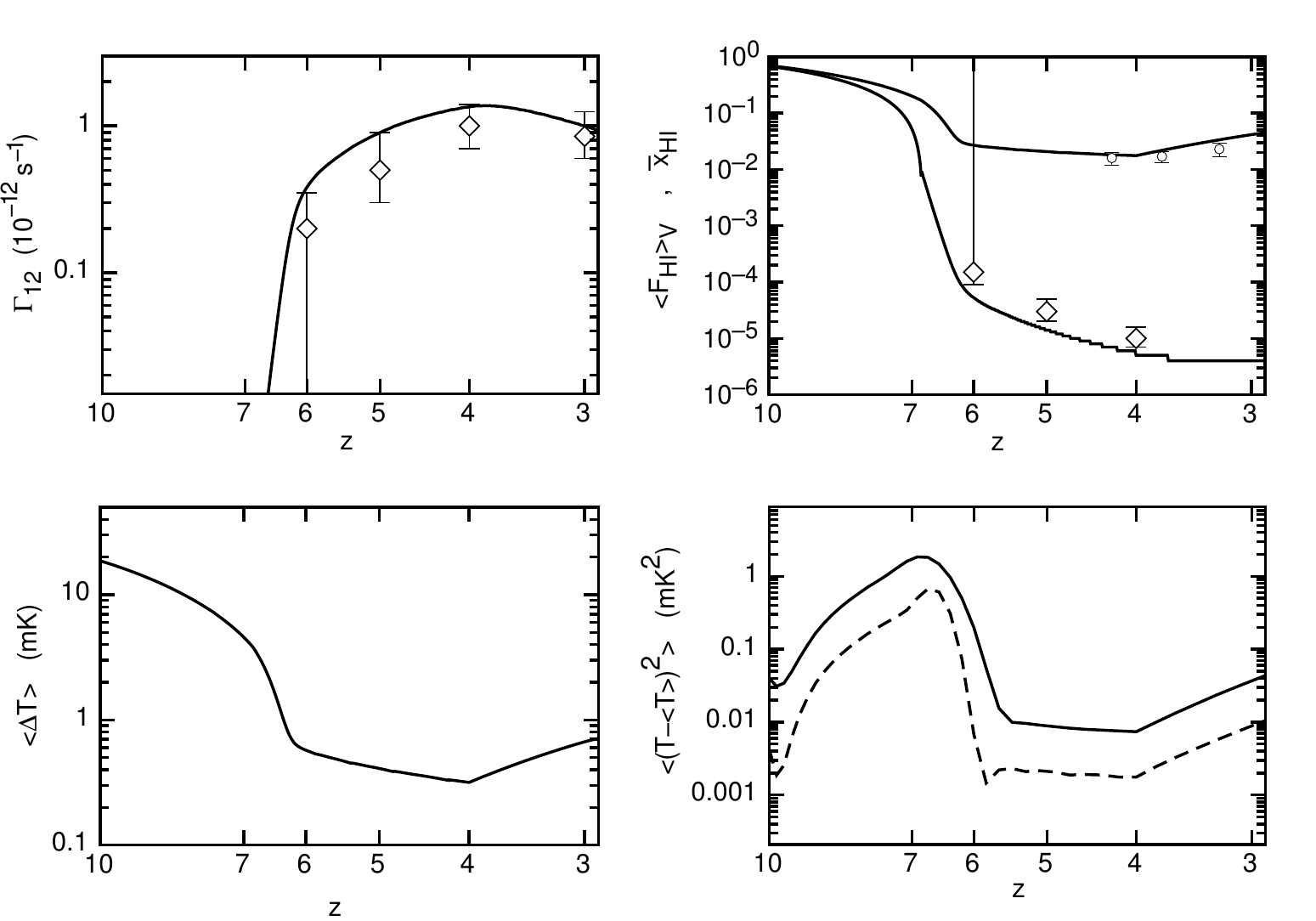}
\caption{Example models for the reionization of the IGM and the subsequent post-overlap evolution of the ionizing radiation field that fit current constraints \citep[based on][]{wyithe2008}. Upper left panel: The ionization rate as a function of redshift. The observational points are from \citet{bolton2007b}. Upper right panel: The volume (lower curves) and mass (upper curves) averaged fractions of neutral gas in the universe. The observational points for the volume-averaged neutral fraction are from \citet{bolton2007b}, while the observed mass fractions are from the damped Ly$\alpha$ measurements of \citet{prochaska2005}. Lower left panel: The predicted evolution of the mean 21~cm brightness temperature (in mK) with redshift. Lower right panel: The auto-correlation function (variance) of brightness temperature plotted as a function of redshift for scales of relevance for observations using upcoming low-frequency radio telescopes (solid and dashed lines assume smoothing scales in real space of 20 and 40 comoving Mpc respectively).}
\label{fig_WL}
\end{center}
\end{figure}

Following the completion of reionization HI was confined
to over-dense regions of the IGM where the enhanced recombination rate
maintains a non-zero fraction of hydrogen in atomic form. The situation therefore differs
between low and high density regions of the universe which contain
optically thin Ly$\alpha$ absorbers and self-shielded damped
Ly$\alpha$ systems (DLAs) respectively. The ionization fraction in the low
density regime is controlled by the balance between the ionization
rate owing to the UV background and the recombination rate at the
local gas density \citep[e.g.][]{bolton2005}.
In contrast, DLAs are known to be dense and
self-shielded \citep{wolfe2005}.  
Although little is understood about the nature of DLAs, it is known that
they play a very important role in the evolution of
hydrogen, containing $\sim80\%$ of the HI in the universe at $z<4$
\citep{prochaska2005}. Their clustering \citep{cooke2006} and
abundance \citep{zwaan2005b}, as well as theoretical expectations
\citep{nagamine2007} suggest that DLAs are housed within galactic mass
systems. 
The upper right hand panel of Figure~\ref{fig_WL} illustrates the mean evolution of the mass and volume weighted HI content of the universe following the end of reionization, compared with predictions of an analytic model. Most of the HI content is in dense clumps, and so does not contribute to the Ly$\alpha$ optical depth.



\subsubsection{The 21~cm brightness temperature}


The 21~cm line, corresponding to the ground state hyperfine transition of atomic hydrogen, has a long history in galactic and extra galactic astronomy. Studies have focused on the inter-stellar medium, either of our own galaxy or of other galaxies. However recent developments in low frequency radio arrays are making it possible to study diffuse HI at high redshift. 

The fluctuations in 21~cm intensity (or brightness temperature) from different regions of the IGM include contributions from a range of different physical properties, including density, velocity gradients, gas temperature, gas spin temperature and ionization state \citep{Furlanetto:2006p341}. 
Quantitatively, the 21~cm
brightness temperature contrast may be written
\begin{equation}
\label{Tb}
\Delta T=23.8\left(\frac{1+z}{10}\right)^\frac{1}{2}\left[1-\bar{x}_i(1+\delta_x)\right]\left(1+\delta\right)\left(1-\delta_v\right)\left[\frac{T_{\rm s}-T_{\rm CMB}}{T_{\rm s}}\right]\,\mbox{mK},
\end{equation}
where $\bar{x}_i$ is the mean ionization fraction ($\bar{x}_i=1-\bar{x}_{\rm HI}$), $\delta$ is the dark matter density fluctuation, $\delta_x$ is the ionization fraction fluctuation, $\delta_v=(1+z)H^{-1}\partial v_r/\partial r$ is the velocity distortion sourced fluctuation (with $\partial
v_r/\partial r$ being the gradient of the peculiar velocity along the line-of-sight with Hubble parameter $H$), and $T_{\rm s}$ and $T_{\rm CMB}$ are the spin temperature of the HI and the temperature of the CMB background radiation respectively. 

The evolution of the ionized fraction $\bar{x}_i(z)$ from zero to one should produce a soft $\sim 20$~mK `step' in the spectrum as a function of frequency towards the end of the EoR~(Figure~\ref{fig_WL}, lower left panel). While the global step \citep[e.g.][]{shaver1999,bowman2008} may provide a mean redshift of reionization and constrain the very high redshift spin temperature evolution \citep{PL08}, observations of  the 21~cm fluctuations (which are comparable in amplitude to the mean signal; lower right panel in Figure~\ref{fig_WL}) have the potential to unravel the processes behind the EoR and probe cosmology. 


In this review we concentrate on the theory and observations associated with 21~cm fluctuations, and restrict our attention predominantly to those predictions for the 21~cm signal of reionization that are most applicable to low frequency arrays which are either in planning or under construction. The fluctuations and evolution of the kinetic and spin temperatures of HI are frontier issues in theoretical studies of the EoR \citep[e.g.][]{PF07,PL08}. However the first generation of EoR instruments will have good signal to noise only during reionization, and as a result we have focused our attention on analyses that assume the spin temperature of hydrogen is coupled to the kinetic temperature of an IGM that has been heated well above the CMB temperature (i.e. $T_{\rm s}\gg T_{\rm CMB}$). This condition should hold during the later stages of the reionization era \citep[$z\lesssim9$][]{S+07} and for collisionally coupled gas
in collapsed objects after the completion of reionization \citep[observed in some DLAs][]{curran2007}. In this regime the term $[(T_{\rm s}-T_{\rm CMB})/T_{\rm s}]$ in equation~(\ref{Tb}) approaches unity and fluctuations in the kinetic and spin temperatures of the gas can be ignored resulting in a proportionality between the ionization fraction and 21~cm intensity. If true, this proportionality greatly simplifies the prediction and interpretation of the 21~cm signal.

\subsection {Forthcoming instruments for 21~cm observations}
\label{Experiments}

While 21~cm EoR and post-reionization cosmology instruments are conceptually quite similar (see \S \ref{ObsSec}), the difference in frequency (redshift)
necessitates that the EoR and lower redshift cosmology measurements will almost certainly be performed by separate instruments. There are four EoR arrays planning to start sensitive observations within the next year and a half, and there are several lower redshift ``intensity mapping'' machines under active development. The HI machines are pushing the state-of-the-art in instrumental and ionospheric calibration, large and ultra-large $N$ correlators, widefield observation, and high dynamic range imaging. In addition to opening a new field of cosmology, these telescopes are laying a technical foundation for future radio instruments at all frequencies. In this section, we briefly introduce the first generation 21~cm fluctuation experiments and discuss their unique characteristics.


\subsubsection{GMRT}
The Giant Metre-wave Radio Telescope\footnote{http://gmrt.ncra.tifr.res.in/} is a general purpose low frequency observatory in the Pune region of India, about 120 km east of Mumbai. The GMRT has thirty 45~m diameter dishes, the central 14 of which are within a 1~km region and are of particular use for EoR power spectrum observations (see \S \ref{HIsensitivity} for a discussion of sensitivity and antenna separations). The GMRT epoch of reionization team has upgraded the digital processing system, developed a novel pulsar-based calibration approach, and developed advanced techniques for identifying and removing both narrowband and broadband radio frequency interference \citep[RFI,][]{Pen:2008p4020}. The GMRT team has already collected several hundred hours of observations, and additional analysis and observations are ongoing.

\subsubsection{PAPER}
The Precision Array to Probe Epoch of Reionization\footnote{http://astro.berkeley.edu/$\sim$dbacker/eor/} is led by researchers at Berkeley, and will consist of 128 antennas located at the Murchison Radio Observatory site in western Australia. The PAPER antennas feature a unique dipole + resonator receiving element and focusing ground plane which are designed to have a very predictable and stable calibration. Unlike the other EoR instruments, the antennas do not track the sky and PAPER operates as a drift scanning instrument. Because PAPER is an EoR only observatory, the antennas can be moved to maximize the tradeoff between sensitivity and instrument calibration (\S \ref{HIsensitivity} and \S \ref{calibration}). Flexibility in design and learning from early observations are distinct strengths of the PAPER experiment.

\subsubsection{LOFAR}
The LOw Frequency ARray\footnote{http://www.lofar.org/} is currently under construction in the Netherlands with antennas extending into several countries of northern Europe.  LOFAR has been at the vanguard of new low frequency observation, and is a multi-purpose observatory with a broad science case including high angular resolution, 10--250~MHz frequency coverage, and high speed cosmic ray detection (see \cite{Rottgering:2006p4017} for overall design and http://www.astron.nl/radio-observatory/astronomers/lofar-astronomers for up to date technical information). LOFAR's EoR observations will use the high band 110--250~MHz antennas in the central two kilometers of the array. Each high band antenna consists of a 4x4 array of dual polarization dipoles on a ground screen and is embedded in a foam environmental protection system. The signals from each dipole can be delayed in an analog beamformer to form a $\sim$20+ degree antenna beam that is electronically steered to track objects across the sky. In LOFAR the individual high band antennas are further grouped into stations of 24 antennas each to form `virtual' $\sim$30~m antennas with multiple beams per station (the number of beams per station depends on available computation, and the more distant stations used for non-EoR observations typically have more antennas per station.) The LOFAR EoR measurement will use the 36 stations near the core of the array (864 antennas) and concentrate on 5 observational windows with low sky temperature and Galactic polarization. LOFAR has the largest collecting area of the widefield EoR instruments (PAPER, LOFAR, MWA) and the highest angular resolution, both of which are important for instrumental calibration and foreground subtraction.

\subsubsection{MWA}
The Murchison Widefield Array\footnote{http://www.MWAtelescope.org/} \citep{Lonsdale:2009p3966} is currently under construction in western Australia at the radio quiet Murchison Radio Observatory (MWA and PAPER are walking distance apart). The MWA antennas are very similar to the LOFAR high band antennas (the telescopes share a technical history), though the MWA antennas are optimized for a slightly higher frequency (140~MHz) and correspondingly have a slightly larger $\sim$30$\degree$ beam. The 512 antennas of the MWA are arranged in a very compact 1.5~km semi-random distribution designed to maximize the quality of the point-spread-function, and features full cross-correlation of all antennas (130,000 baselines in 4 polarizations and 3000 frequency channels) and realtime holographic calibration. The full correlation allows the entire antenna field of view to be imaged, enhancing the survey speed of the instrument. Of the first generation EoR instruments, the MWA has the highest power spectrum sensitivity (by a small margin) and concentrates on very low systematics in its calibration and foreground subtraction.

\subsubsection{21~cm intensity mapping}
\label{ct}

Over the last couple of years there has been significant interest in studying cosmology via 21~cm fluctuations after reionization with EoR-style instrumentation.


It has been recognized for some time that the Square Kilometer Array (SKA) could perform a billion galaxy spectroscopic redshift survey in HI \citep{Rawlings:2004p1676}, and that such a survey could have very low systematic errors for cosmology (see the report of the Dark Energy Task Force\footnote{http://www.nsf.gov/mps/ast/detf.jsp}). The proposed SKA BAO dark energy measurement would identify the three dimensional positions of bright galaxies ($\gtrsim~L_{*}$), and use them as tracers of the underlying density fluctuations. The imprint of baryon acoustic oscillations on the galaxy clustering provides a standard ruler, enabling the expansion history of space to be determined out to a redshift of $\sim$1.5.  In addition to precision dark energy constraints, an SKA HI survey would provide a wealth of astrophysical data about galaxy formation and evolution.


\citet{WLG08,chang2008} and \citet{Peterson:2006p3971} realized that compact interferometers with wide fields-of-view might be able to perform cosmological measurements, particularly those associated with the BAO scale, at a fraction of the SKA's cost. These proposed instruments have low angular resolution---they cannot resolve individual galaxies---and have been tuned to have the highest sensitivity at BAO angular scales ($\sim$150~cMpc). Somewhat counter-intuitively the low angular resolution \emph{enhances} the cosmological sensitivity of these instruments in two ways. First, by observing the integrated 21~cm emission in $\sim$150~cMpc regions they observe all of the HI emission from the region, not just the emission from the brightest galaxies in that volume. Effectively the observation integrates the full luminosity function, and there is a lot of HI in DLAs and other small galaxies. Secondly, by moving the antennas to short spacings all of the baselines contribute to the measurement. 
This observational technique is commonly called `redshifted 21~cm intensity mapping' \citep{Peterson:2009p3975} in reference to the focus on large scale 21~cm birghtness fluctuations, or just `intensity mapping' for short.



Current instrument concepts for 21~cm intensity mapping include CHIME (Canadian Hydrogen Intensity Mapping Experiment) and the related CRT (Cylinder Radio Telescope), the Omniscope (formerly FFT Telescope), and CARPE\footnote{http://www.phys.washington.edu/users/mmorales/carpe/} (Cosmological Acceleration and Radio Pulsar Experiment) telescope. Several of the arrays are being prototyped, though none have construction funding as of publication.

All of the proposed 21~cm intensity mapping instruments have wide fields-of-view to maximize the survey speed (100's of square degrees), and thousands of receiving elements. The large number of receiving elements presents a significant data processing challenge because radio correlators scale with the number of receiving elements \emph{squared}.  At 512 elements the MWA correlator is the largest $N$ correlator ever constructed, and even Moore's law does not erase the problem for arrays contemplating an order-of-magnitude more receiving elements. The current instrument concepts are largely driven by data processing considerations, using either cylinders (CHIME and CRT), a regular grid of elements and FFT algorithms (Omniscope), or novel correlator designs (CARPE). We expect that instrumental calibration will become an important design driver as these concepts mature, and we hope at least one first generation 21~cm intensity mapping experiment will be under construction within a few years.

In this review we have integrated the discussions of 21~cm EoR and post-reionization intensity mapping measurements to make the connections between the fields explicit. Section \ref{signal} reviews the theoretical expectations for 21~cm emission with an emphasis on advances made in the past three years. Section \ref{ObsSec} then concentrates on upcoming observations, with particular attention to the cosmological sensitivity of interferometers, instrumental calibration, and foreground subtraction.


\section{Evolution of the ionization structure in the IGM}

\label{signal}

As detailed in \S~\ref{overview}, present day constraints on the reionization of hydrogen are limited to the evolution of the mean ionized fraction (see Figure~\ref{fig_WL}). However within our modern theoretical framework it has been recognized that reionization is not a uniform process. Rather, hydrogen reionization is thought to have started with small ionized (HII) regions around the first galaxies to form, and that these later grew to surround groups of galaxies. Reionization is said to have completed once HII
regions overlap, occupying most of the inter-galactic volume and leaving only dense islands of neutral gas (as shown in Figure~\ref{Trac1}).  Thus the process of reionization is thought to be inhomogeneous, owing to both fluctuations in the density and to fluctuations in the ionization structure of the IGM. The inhomogeneities during and after reionization source 21~cm intensity fluctuations, facilitating the study of both galaxy formation and cosmology. 

In this section we review current theoretical expectations for the HI signal, concentrating on those that can be studied with upcoming observations. Sections \ref{PSsec}--\ref{EoRCosmo} focus on the signal during reionization, which is complicated by the source properties of the first luminous objects, the impact of reionization and heating on formation of the subsequent generations of objects, radiative transfer, and other effects. Section \ref{aftereor} then considers the simpler post-reionization signal which is important for cosmological measurements and astrophysics at more recent times.

\subsection{The power spectrum}
\label{PSsec}

The statistics of the 21~cm intensity field resulting from the reionization process are often described using the power spectrum of 21~cm fluctuations (hereafter referred to as the 21~cm PS). Measurement of this PS is one of the primary goals for low frequency radio interferometers. Before describing the modeling efforts that have been undertaken to
predict, and eventually interpret 21~cm observations, 
it is therefore useful to first introduce the PS of 21~cm emission arising from
ionization structure in the IGM. 

The primordial density field as revealed by the CMB may be described as a Gaussian
random field, in which the distribution of over-densities in
the Fourier decomposition of the field is Gaussian at all spatial
frequencies.
This scale dependent variance is referred to as the PS. In the
absence of ionizing sources and on large scales, the PS of
HI follows the dark matter PS.
However once the first ionizing sources turn on, the evolution of
ionized structure evolves in a complex way that reflects the
astrophysics of the ionizing sources as well as the cosmology of the
underlying density field.  

To leading order in $\delta$, it
follows from equation~(\ref{Tb}) with $T_{\rm s}\gg T_{\rm CMB}$ that the PS of brightness temperature
fluctuations is given by
\begin{eqnarray}
\nonumber
\label{P21}
P_{\Delta T}&=&\mathcal{T}_{\rm b}^2\left[\left(\bar{x}_{\rm HI}^2P_{\delta\delta} - 2\bar{x}_{\rm HI}(1-\bar{x}_{\rm HI}) P_{\delta x} + (1-\bar{x}_{\rm HI})^2P_{xx}\right) \right.\\
&&\hspace{-10mm}+2\f\mu^2\left(\bar{x}_{\rm HI}^2P_{\delta\delta}-\bar{x}_{\rm HI}(1-\bar{x}_{\rm HI})P_{\delta x}\right)+\left.\f^2\mu^4\left(\bar{x}_{\rm HI}^2P_{\delta\delta}\right)\right],
\end{eqnarray}
where $\mathcal{T}_{\rm
  b}=23.8\left[{(1+z)}/{10}\right]^\frac{1}{2}\,$mK, and
$P_{\delta\delta}$, $P_{\delta x}$ and $P_{xx}$ are the PS of density
fluctuations, the cross-PS of ionization and density fluctuations, and the PS of ionization fluctuations respectively \citep{mao2008}.  In equation~(\ref{P21}) the velocity fluctuation has been written (to lowest order in Fourier space), as
$\delta_v(\vec{k})=-\f\mu^2\delta$ where $\mu$ is the cosine of the angle
between the $\vec{k}$ and line-of-sight unit vectors~\citep{kaiser1987},
and\footnote{The quantity $\f$ is close to unity at high redshifts, taking
values of 0.974, 0.988 and 0.997 at $z=2.5$, 3.5 and 5.5.}
$\f=d\log{\delta}/d\log{(1+z)}$. 
As with traditional galaxy redshift surveys the term $P_{\delta\delta}$ encodes
the cosmology. However during the EoR the terms $P_{\delta x}$ and $P_{xx}$ are sensitive to
the astrophysics of the reionization process, and represent the
cross-correlation between the clustering of galaxies and the resulting ionized structure, and the
auto-correlation of ionized regions respectively.

Late in the reionization process the terms $P_{\delta x}$ and $P_{xx}$ dominate 21~cm  fluctuations, and so the formation of large HII regions has a significant effect on the shape of the 21~cm PS. This effect arises for two reasons. First, information about
small scale features in the density field is erased from the signal
originating within the HII regions. In addition, the creation of
HII regions imprints large scale fluctuations on the
distribution of 21~cm intensity. The sum of these effects is to move
power from small to large scales, leaving a shoulder shaped feature on
the PS at the characteristic scale of the HII regions
that moves to larger scales as the bubbles expand during the latter stages of reionization \citep{furl2004a} . This evolution is illustrated by the sequence of power spectra shown in Figure~\ref{Barkana2008Fig4}. Unlike CMB PS measurements, HI EoR measurements can observe a series of redshifts, creating a coarse power spectrum `movie' of reionization. The dynamics of the PS as reionization progresses provide crucial information for constraining models of reionization \citep{lidz2008}. 



\subsection{Modeling the ionization structure to predict the 21~cm signal}
\label{modeling}

In anticipation of forthcoming 21~cm EoR observations, a great deal of
theoretical attention has focused on the prospects of measuring the 21~cm PS. 
This theoretical work can be divided into analytic studies, full numerical simulations, and semi-numerical approaches.

\subsubsection{Analytic models}


We begin by discussing analytic descriptions of the reionization process. 
A very successful analytic model describing the statistics of
ionization structure during hydrogen reionization (the bubble model) was introduced by
\citet{furl2004a,furl2004b}. This model computes the mass function of
HII regions using an analogy of the excursion set formalism for
computing the dark matter halo mass function. The crucial difference
between the bubble model formalism and the standard excursion set formalism
is that the barrier is a (decreasing) function of HII region 
size, instead of a constant. This feature gives HII regions a characteristic
size rather than the power-law behavior seen at small masses in the
\citet{ps1974} mass function. The characteristic scale \citep[which is much larger than expected for an individual star forming galaxy,][]{wl05} is found to
depend primarily on the ionization fraction.


The bubble model is based on one point statistics and naturally produces the probability distribution for
sizes of HII regions. However in order to compute the 21~cm PS 
the bubble model employs approximate expressions for the spatial
correlation of ionizations \citep{furl2004a} which are grafted externally onto the underlying formalism. To overcome this, \citet{B07} found a solution
for 2 correlated random walks, and calculated the 2-point correlation
function (and hence PS) directly. 
As discussed in \citet{barkana2008}, the analytic and numerical calculations do not agree in detail. On the other hand numerical simulations do not yet agree among
themselves at a comparable level \citep{ilievc2006}. 

Modeling of the 21~cm PS (Figure~\ref{Barkana2008Fig4}) indicates 
that the 21~cm PS has features that vary
non-monotonically, providing opportunities for a range of parameters
to be measured. Owing to the computational efficiency of the extended bubble model, \citet{barkana2008} argued that observations of the evolving 21~cm
PS could therefore be used to constrain critical physical parameters
in the reionization process.
To achieve this \citet{barkana2008} suggested that the analytic model could be tuned to provide a sufficiently precise
description through comparison with numerical simulation (with the caveat that the simulations have converged to a reliable prediction). 
Under this assumption, a fast analytic model could play
the same role with respect to identifying reionization parameters from
observations of a 21~cm PS that the code CMBfast does for the
CMB. However in addition to issues of precision, it should be noted
that analytic models for the 21~cm PS do not yet include all possibilities, either for the ionizing
sources or the contributions to the 21~cm signal which could lead to
additional systematic uncertainty in inferences of source population
from the 21~cm observations. We return to discuss this point further in
\S~\ref{sec_contributions}.

On scales much larger than the bubble size analytic models have also been used to investigate the probability distribution for 21~cm brightness temperature.
Galaxy bias is found to lead to enhanced
reionisation in over-dense regions. Moreover, because galaxy bias operates most strongly on the
exponential tail of the \citet{ps1974} mass function the
resulting enhancement is not linear with over-density, leading to a non-Gaussian distribution of 21~cm
brightness temperature in intensity maps of reionisation. 
\citet{WM07} demonstrate that skewness of the intensity distribution is present on all
scales, and during the entire reionisation era. This large scale skewness is related to, but is different from from the excess power (or shoulder) induced in the
PS at small scales following the appearance of HII regions
\citep{furl2004a}. Importantly, neither the skewness
nor the variance (calculated within real-space volumes) are monotonic with redshift or neutral fraction, even in a simple
reionisation scenario.

The sum of these analytic results describes the following fundamental features that form the basis of our modern picture for the structure of the reionization process:
\begin{itemize}
\item Large-scale,
over-dense regions near sources are generally reionized first, while
the under-dense regions are reionized via the overlap of neighbouring
HII regions near the end of the reionization process. 
\item Sources which are massive tend to be surrounded by an
over-density of more numerous sources owing to galaxy
clustering. These increase the size of HII regions
beyond what would be expected from the massive source, and can dominate the ionizing contribution to some HII regions.
\item Non-Gaussianity is a generic feature of 21~cm fluctuations during the EoR, both on scales that are comparable to, and much larger than individual HII regions.
\end{itemize}

\subsubsection{Numerical Models}
\label{numerical}

While the analytic studies described above reveal many of the generic features of the reionization process, the inhomogeneous, non-linear and non-Gaussian nature of reionization ensures that detailed modeling and understanding of the ionization structure requires numerical simulation. There
are a range of different techniques applicable.  Solving the reionization
problem requires N-body techniques to calculate the evolution of dark
matter, hydro-dynamical techniques to model the gaseous component, and
an algorithm for radiative transfer to solve evolution in
ionized structure. The challenge for modeling of the reionization
process via direct simulation is to resolve all of the spatial scales
involved. In particular, high resolution is required to resolve small
scale structure such as galaxies, as well as the Ly-limit systems that regulate
radiation propagation and so form one of the key factors regulating
reionization. At the same time, large simulation volumes are necessary
in order to contain a representative distribution of
galaxies and HII regions. Illustrating the difficulty of these combined requirements is that both analytic models and simulations have shown 
HII regions to reach characteristic sizes (10's of Mpc) that contain a mass
7 orders of magnitude larger than the halos with a virial
temperature of $T=10^4$K that are thought to drive reionization. 

Experience has shown that a realistic study of the large-scale properties of reionization requires a minimum box size of around $100$ co-moving Mpc on a side. This minimum size is set by 2 independent requirements.  First, as mentioned above the typical HII region
size is found to be tens of co-moving Mpc during much of the EoR~\citep{furl2004a}. If one wants
to study the structure and statistics of HII regions during
the EoR, the simulation must contain many of these HII regions. The
second reason concerns cosmic variance. As pointed out by
\citet{barkana2004} a simulation which is too small does not contain
the large scale density fluctuation modes which drive the clustering of galaxies on the largest scales. An undersized simulation will predict an unphysically rapid reionization, owing to 
periodic boundary conditions which remove the corresponding large scale
fluctuations in the radiation field. 

Over this past decade, significant progress has been made in modeling the effect of galaxies on
the reionization of the IGM. Pioneering studies
\citep[e.g.][]{ciardi2003,sokasian2003} were restricted to small box
sizes of only $\sim10 - 20$ Mpc, with a consequently small number of
sources. However the results of these works illustrated that the
abundance and luminosities of galaxies clearly affect the reionization
process. In large modern simulations, the most common approach to
address this issue is to begin with an N-body code to generate a distribution of
halos \citep[e.g.][]{ciardi2003,sokasian2003,iliev2007,zahn2007,trac2007}. A simple prescription is then used to relate halo mass to
ionizing luminosity. Following this step, radiative transfer methods (most commonly ray-tracing algorithms) are employed to model the generation of ionized structure on large scales. The radiative transfer is normally
run with lower resolution than the N-body code for computational
efficiency.

In the last few years, the critical box size of $100 h^{-1}$ Mpc for
reionization simulations has been reached in this
way~\citep[e.g.][]{iliev2007,zahn2007,mcquinn2007}. More importantly, in just the last year simulations have become
sufficiently large that they have met both the box size and mass
resolution requirements \citep{shin2008,Il08,trac2008}. In
Figure~\ref{Trac1} we reproduce a sample visualization of the large
scale simulation presented in \citet{trac2007}. Simulations such as
this describe the generic features of
reionization
\citep[e.g.][]{iliev2007,zahn2007,mcquinn2007,shin2008,lee2008,croft2008}, confirming expectations from analytic models that large-scale,
over-dense regions near sources are generally reionized first, and that massive galaxies tend to be surrounded by clustered sources that increase the size of HII regions. In addition, the simulations describe the structure of the HII regions, showing that they are generally aspherical (even where the sources are assumed to emit isotropically). 

The evolution of ionized structure is affected by the presence of
Ly-limit systems which absorb the ionizing photons before they can
contribute to ionization of the diffuse IGM. This effect has been studied by
\citet{mcquinn2007} and \citet{croft2008} using a range of simulations, and
considering different models for both the generation and absorption if
ionizing photons. These studies show that the size and structure of ionized bubbles at fixed global neutral fraction is more strongly influenced by ionizing source properties than by the presence of Ly-limit systems, whose impact is to slow the progression of reionization. The growth of HII regions during reionisation may also be influenced by radiative feedback in
the form of suppression of galaxy formation below the cosmological
Jeans mass within a heated IGM
\citep{dijkstra2004}, although the importance of this effect remains controversial \citep{mesinger2008}. Suppression of low mass galaxy-formation delays and extends the reionization process, which though started by low mass galaxies, must then be completed by relatively massive galaxies~\citep{iliev2007}. 
Conversely, ionization also reduces the clumping factor of the IGM
\citep{pawlik2009} by increasing
the pressure support within dense systems.
The balance of these effects is critical for understanding the detailed process of reionization.

An important outcome from the large cosmological volumes attained by 
modern numerical simulations has been the prediction of 21~cm
signals that will be observable using forthcoming low frequency arrays
\cite[e.g.][]{mellema2006,lidz2008}. The most generic features of 21~cm PS modeling 
were elucidated by \citet{lidz2008}, who show examples of its evolution. On scales of $k\sim0.1$Mpc$^{-1}$ the terms $P_{\delta x}$ and $P_{xx}$ modify both the amplitude and the slope of the PS in a non-monotonic way relative to the expected shape in the absence of ionization structure (see Figure~\ref{lidzfig2}). 
Thus, measurement of these terms will provide the first clues regarding the clustering of
ionizing sources during the EoR. \citet{lidz2008} illustrate that
the slope and amplitude of the 21~cm PS vary considerably
among different models at a given ionization fraction. However
they also find that the behavior with ionization fraction across the
different models is relatively generic. In particular, the amplitude of the 21~cm
PS reaches a maximum close to the epoch when $\sim 50\%$
of the volume of the IGM is ionized, while its slope is found to flatten with
increasing ionization fraction as seen in Figures \ref{Barkana2008Fig4} and \ref{lidzfig2}. \citet{lidz2008} argue that a
first generation EoR telescope like the MWA has sufficient sensitivity to measure the
redshift evolution in the slope and amplitude of the 21~cm power
spectrum. Moreover the characteristic redshift evolution evident in
models of reionization that are driven by star formation will allow
the mean ionization fraction near the center of the EoR to
be constrained to a level of $\sim10\%$.

Finally, in addition to the full 3 dimensional structure of ionization during the EoR, numerical simulations are able to follow the thermal
properties of the IGM. The simulations shown in Figure~\ref{Trac1} illustrate that the IGM
in over-dense regions is photo-heated earlier than large-scale,
under-dense regions far from sources. 
Thus, conditions in the Ly$\alpha$ forest at $z<6$ can be used to
probe the ionization state during the EoR and to tell the
difference between different scenarios for the reionization history~\citep{trac2008}.

\subsubsection{Semi-numerical models}

In order to overcome the limitations of both analytic and numerical
methods \citet{MF07} introduced an approximate, but efficient method for
simulating the reionization process. This so-called {\em semi-numerical} method extends prior
work by \citet{bond1996a} and \citet{zahn2007}, and has 2 parts. First, an excursion-set approach
\citep[e.g.][]{lacey1993} is applied to identify halos within a
realization of the linear density field, following which halo
locations are adjusted with first-order perturbation theory. The
method therefore generates halo distributions at a particular
redshift, without explicitly including information from higher
redshift. 
Despite this simplification, the mass and correlation functions of the
resulting halo population 
agree almost perfectly with results from N-body simulation.

\begin{figure}
\begin{center}
\includegraphics[width = 5.25 in]{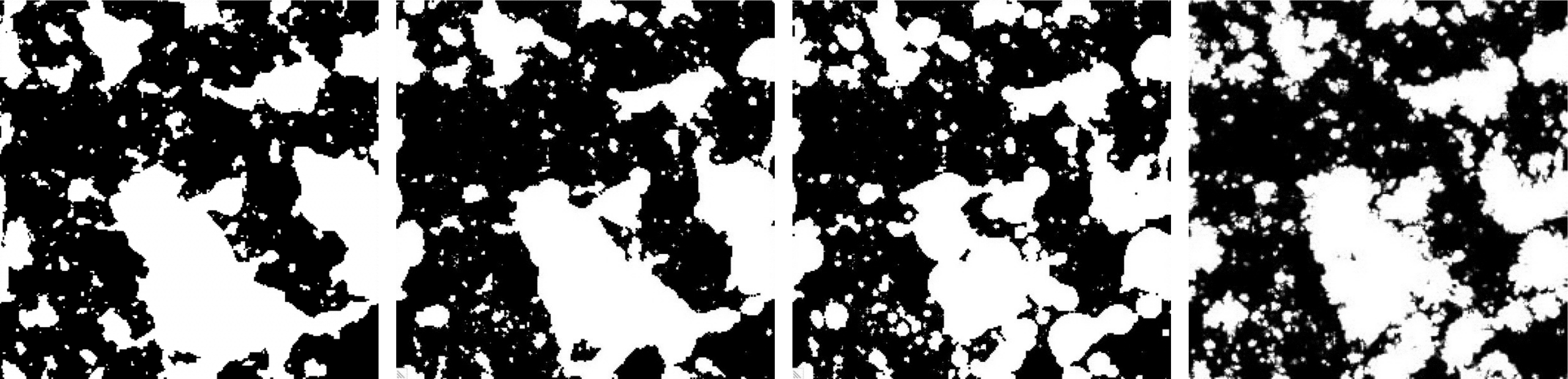}
\caption{From \citet{MF07}. Slices from the ionization field at $z = 6.89$ created using different algorithms. All slices are 93.7 Mpc on a side and 0.37 Mpc deep, with the mean neutral fraction in the box being $\bar{x}_{\rm HI} = 0.49$. Ionized regions are shown as white. The left most panel was created by performing the bubble-filtering procedure of \citet{zahn2007} directly on a linear density field. The second panel was created by performing the bubble-filtering procedure of \citet{zahn2007} on their N-body halo field. The third panel was created by performing the bubble-filtering procedure of \citet{MF07} on the same N-body halo field. The rightmost panel \citep[from][]{zahn2007} was created using a ray tracing algorithm on the same halo field.}
\label{MessingerFurlanetto2}
\end{center}
\end{figure}

\citet{MF07} then estimate the ionization field based on a catalogue of sources assigned within the halo field \citep[an idea previously used by][]{zahn2007} by applying a filtering technique based on the analytic bubble model. 
Figure~\ref{MessingerFurlanetto2} illustrates the structure of the
ionization field obtained, including comparison of different variations on the
filtering method and full
radiative transfer. There is good agreement, implying that semi-numerical models can be used to explore a larger range of reionization
scenarios than is possible with current
numerical simulations.  Importantly, these semi-numerical models retain
information on the spatial distribution of sources and ionization
structure that are not available via purely analytic modeling.

\citet{thomas2009} introduced a related technique in which a large N-body simulation is used as the basis for construction of the ionization field in post-processing. In this scheme sources are again assigned to the halos with appropriate spectral properties. However rather than adopt a filtering scheme, a library of pre-computed 1-dimensional radiative transfer results are used to generate HII regions based on the spherically averaged radial density profile. A feature of this model is that overlap of neighbouring HII regions is treated self consistently with respect to photon conservation (photon conservation is not guaranteed within the filter based semi-numerical methods described above).

Recent work by \citet{choudhury2009} provides an excellent example of the utility of semi-numerical modeling. These authors looked at the effect of a position dependent clumping factor on the
structure of reionization. Their main point of departure from previous
work is the enforcement of consistency with the Ly$\alpha$ forest
data, which leads to a {\em photon starved} reionization
\citep{bolton2007b}. In the photon-starved regime,
\citet{choudhury2009} find that recombinations are much more important
than in models where reionization occurs quickly, and that accounting
for spatially inhomogeneous sinks of ionizing radiation has a large
effect on the topology in this case. Reionization is found to
proceed inside-out during the early stages. However in the photon
starved regime the sinks of ionizing radiation can remain neutral so
that reionization proceeds deep into the under-dense regions prior to
the completion of overlap.  

\subsection{Comparison of modeling methods for the EoR}

Given the various complementary methods in use for simulation of the EoR and the impending comparison between theory and observation, it is timely to compare the theoretical situation with the related case of galaxy clustering and large-scale-structure. Galaxy redshift surveys have reached a point of observational maturity where full N-body simulations of structure formation are required to
provide a sufficiently precise theoretical model. Given the correct
cosmological model, N-body simulations are able to reproduce the details of galaxy clustering because the source of structure in the
density field, namely gravity, is well understood. However both
simulations and observations of the the EoR are currently in a
different position. Firstly, for the foreseeable future 21~cm based observational constraints on the EoR will be crude by comparison with modern galaxy redshift surveys. Moreover, the numerical problem is less precisely posed than a calculation of
structure formation, because the source of structure during the EoR is provided by radiation from galaxies
(which act via radiative transfer), rather than by gravity. However in difference to the structure formation
case where gravity is not generally the topic of study, understanding the
galaxies which drive reionization (i.e. the source of structure) represents the primary scientific goal for studies of the EoR. Currently, the properties of these early galaxies are
very uncertain even as regards their qualitative features (e.g. mass, stellar initial mass function,
contributions from UV photons verses X-rays etc.).  Models of
the EoR that employ different assumptions are therefore able to predict a wide range of
scenarios. Indeed, the range of predictions for quantities like the 21~cm PS is much larger than the uncertainty in
calculations of a particular model, and is also larger than the difference
between different modeling techniques. 

Different methods for performing calculations of the reionization process each have their own advantages and disadvantages, and so it is important to compare predictions where possible. Analytic models utilize the \citet{ps1974} formalism (with extensions) and operate under the premise that the production of ionizing photons is proportional to the formation of collapsed, bound structures \citep{BL01}. It must therefore be recognised that while analytic models are based around a linear theory of structure formation, both the collapsed structures and the response of the IGM to the ionization field are explicitly non-linear. As a result, it is crucial that conclusions reached via analytic studies are checked against
simulations. Never-the-less, analytic models are able to describe the primary features of the reionization process from a fundamental point of view. Moreover, as described above,
simulations with sufficient volume to accurately represent the process of reionization have only recently been performed due to the enormous dynamic range the simulations must
address. The large amount of computation limits the number of model parameters which can be explored. To overcome this limitation it is often advantageous to utilize analytic models, which are able to provide a fast exploration of parameter space in many of the unknown quantities describing early galaxy formation.

\citet{S+07} have compared the results of analytic models to numerical
simulations, both in the high redshift regime where the spin
temperature is an important consideration, and at lower redshifts
where they find that it is not important ($T_{\rm s}\gg T_{\rm CMB}$). 
Moreover at the redshift range that
will be probed by planned 21~cm experiments (i.e. $z\lesssim9$),
\citet{S+07} find that the simple analytical models described above
coupled with fitting functions to describe the ionization fraction
PS are able to reproduce the results from numerical
simulation. The implication is that at these redshifts, there are no
impediments regarding the use of an analytical (or semi-numerical) model to explore the
parameter space relevant for future EoR observations (at least when
restricted to using estimators based on the 21~cm PS alone.)

\subsection{Additional contributions to ionization and the power spectrum}
\label{sec_contributions}

The modeling of the ionization structure and 21~cm PS
discussed thus far has centered around the assumption that the
ionizing photons are dominated by UV flux from star-forming
galaxies. In this section we discuss some possible scenarios which
complicate this picture, and hence threaten to muddy the interpretation of 21~cm
observations. We again restrict our attention to the domain late in
the EoR when the spin temperature of the HI is thought to be much
larger than the CMB \citep[i.e. $z\lesssim9$][]{S+07}. 

\subsubsection{X-rays}
	
The HII region dominated ionization structure of the IGM seen in numerical
simulations (e.g. Figure~\ref{Trac1}) forms as a result of the biased clustering of sources,
combined with the short mean free path (MFP) of UV photons. 
However in addition to UV photons from the first galaxies and quasars, it has also 
been suggested that a background of X-ray photons may have been produced at high
redshift, primarily from black hole accretion.
Unlike UV photons, X-rays ionize hydrogen both directly and through
secondary ionizations by photo-electrons from ionized helium, with the
latter dominating so that many hydrogen atoms can be ionized by a
single X-ray photon. For ionization fractions $\gtrsim 10\%$,
ionization by secondary electrons becomes inefficient \citep{RO04a}
and X-ray ionization is self regulating at the level of $\sim10-20\%$.
Observational limits on the contribution of X-rays to reionization are
provided by the unresolved soft X-ray background, which limits the
possible level of ionization by X-rays from a population of black holes at
high redshift \citep{D+04}.  

		
Importantly, X-rays also 
differ from UV photons in their effect on the IGM due to their
comparatively long MFP, which means that X-rays provide a weakly
fluctuating radiation background.
For this reason, the inclusion of an X-ray component of the
emission responsible for the reionization of the IGM could modify
fluctuations in the 21~cm PS by reducing the influence of strongly
biased ionized regions \citep{warszawski2009}. 
The details of the 21~cm 
PS shape are found to depend on the typical X-ray photon energy, while the
magnitude of modification to the PS is determined by the relative
contribution of X-rays to the EoR. Thus the possible contribution of
X-rays has the potential to substantially complicate
analysis of 21~cm PS observations. 

\subsubsection{Quasars}

Luminous quasars
are known to exist at the edge of the reionisation epoch
\citep{F06}, although the exact role that they play in the reionisation process remains uncertain.  It is thought that the observed quasar population could not have supplied
sufficient UV photons to reionize the Universe at $z>6$
\citep[e.g.][]{loeb2009}. 
On the other hand, high-redshift quasars have a large
clustering bias \citep{Shen2007}, implying that the effect of their
ionising contribution on the ionization PS 
may be significant during the latter part of the
pre-overlap era. Indeed, \citet{geil2009} have shown that with only a $10\%$ contribution to reionization, the cumulative ionising effect of
quasars could result in a scale dependent suppression of the 21~cm PS by up to 30 per cent at constant global neutral fraction. 

\subsubsection{Galactic HI}

In the modeling described in \S~\ref{modeling} only the component of
HI residing in the IGM was considered in forecasting
the statistical 21~cm signal. However, after the completion of
reionisation there is known to be a residual HI fraction of a few
percent in high density clumps which are believed to reside within
galaxies \citep[e.g.][]{prochaska2005}, and which are identified as DLAs. 
This high density, galactic contribution to the HI
content of the Universe must also be present during the reionisation era,
and should therefore be included in predictions of 21~cm fluctuations.
\citet{wyithe2009} investigated the impact of HI in galaxies on the
statistics of 21~cm fluctuations using semi-numerical models.  Assuming that
 2\% of hydrogen is in the form of HI and
located within galaxies during the reionisation era (approximately the mass weighted fraction in the range  $1< z<5$), a scale dependent reduction of 10-20\% is found in the amplitude of 21~cm fluctuations. 
Moreover, because the galactic
HI is biased towards HII regions, inclusion of galactic HI is found to decrease the prominence of the HII region induced {\em shoulder} in the 21~cm PS.

\subsection{Other 21~cm signatures of reionization}
\label{otherSig}


Thus far we have concentrated this review on
observation of the PS of 21~cm fluctuations together with its
evolution with redshift. However there are several other observational signatures that have been suggested. 

\subsubsection{Cross-correlation of galaxies with the 21~cm signal}

\label{crosscorr}

The cross-correlation of 21~cm emission with galaxies would directly probe
the connection between reionization and the sources of ionizing
radiation \citep[e.g.][]{WL07,furlanetto2007,lidz2009}.
For example, 
the generic model prediction that
over-dense regions will be ionized early (see \S~\ref{modeling}) leads to an anti-correlation between 21~cm emission
and the galaxy population. \citet{WL07}  argue that the first surveys should be
sensitive enough to tell the difference between ``outside in''
verses ``inside-out'' reionization, even when combined with galaxy
surveys of only several square degrees.

\citet{mcquinn2007} have considered the specific case of a
cross-correlation of Ly$\alpha$ emitters with 21~cm emission. Their
numerical simulations include the important effects of IGM opacity to
the Ly$\alpha$ line.
They find that the large
HII regions formed during the EoR modulate the observed
distribution of Ly$\alpha$ emitters and boost their observed
angular clustering.  
Subsequently,
\citet{iliev2008} have also studied the cross-correlation of
Ly$\alpha$ emitters with 21~cm emission using a large numerical
simulation. In difference to \citet{mcquinn2007} these authors find a
minimal effect of patchy reionization on the clustering of Ly$\alpha$
emitters, which may be attributed to the large scatter in opacity
along different sight lines, or to uncertainties in modeling of
IGM opacity \citep{dijkstra2007}.

		
\citet{furlanetto2007} discuss the properties of cross-correlation between galaxies and fluctuations in the 21~cm
signal more generally, and outline the main points of possible advantage relative to a PS analysis \citep[see also][for a discussion of the cross-correlation  between 21~cm
emission and the IR background or Sunyaev-Zel'dovich effects]{slosar2007}. They find that the signal-to-noise ratio achieved would exceed
that of the 21~cm PS by a factor of several, 
provided that a suitable galaxy survey is available over the large
field of view (100's of square degrees). Moreover the cross-correlation would allow more
efficient detection of inhomogeneous reionization. 
Finally, detection of a
cross-correlation function would also unambiguously identify the
cosmological nature of the 21~cm fluctuations. 


\subsubsection{Imaging of quasar HI regions}
	 	 
The most striking example of a cross-correlation between the 21~cm signal
and astrophysical sources will be provided by the imaging of quasar HII regions. As mentioned in \S~\ref{overview}, several
of the most distant known quasars exhibit evidence for the presence of an HII region in a partially neutral IGM \citep[e.g.][]{cen2000}, although the interpretation remains uncertain
\citep{lidz2007, bolton2007}.
A better probe of the neutral IGM in the quasar environment during the EoR
will be provided by redshifted 21~cm emission, particularly since
quasar HII regions are predicted to be the most prominent individual
21~cm features \citep{kohler2005}.  It has been argued that these
signatures will be most readily detected a-posteriori, around known
high redshift quasars \citep{GWPO08}. Indeed, patchy reionization renders blind detection of HII regions almost impossible for
neutral fractions smaller than about 60\% \citep{datta2008}. \citet{valdes2006} and
\citet{GWPO08} have studied the impact of a percolating IGM on the
detection of HII regions around known quasars, and shown that large (many arcminute) HII regions will provide a detectable imprint on the IGM until very late
in the reionisation era, provided that their location is known a-priori. 
As an example, slices through a model of 21~cm emission from the IGM
surrounding a high redshift quasar are presented in the upper row of Figure~\ref{Wyithe1}.  
The evolution of the IGM is clearly seen in this figure, with the
percolation process completing between the ``back" of the box and the ``front" of the box ($x_3$ coordinate, corresponding to the line-of-sight direction).

Quasars have a harder spectrum than stars, and as pointed out by
\citet{zaroubi2005} and \citet{kramer2008}, this leads to thicker
ionizing fronts than is the case for a starburst driven HII
region. These authors argue that with sufficient angular resolution
the contribution of hard ionizing sources such as mini quasars to
the EoR could be inferred from the structure of ionizing fronts
at the edge of HII regions. 
Moreover, quasar HII regions will last for a long time following the epoch when the
quasar turns off.
Fossils of quasar HII regions in an inhomogeneous IGM at $z<10$ would maintain an ionization state greater than
$\sim90\%$ for around a Hubble time following the death of their
quasar~\citep{furlanetto2008}. While imaging of the resultant networks of HII regions (as opposed to a single quasar HII region) is beyond the reach of the first generation EoR observatories (\S \ref{HIsensitivity}), this will be a primary goal of second generation of EoR telescopes.


\subsubsection{PDF of 21~cm intensity fluctuations}

The distribution of fluctuation amplitudes is expected to be Gaussian
early in the EoR. At this time the PS is the natural statistic to 
describe fluctuations in 21~cm emission, since it contains all the statistical
information. However the distribution of fluctuation amplitudes becomes non-Gaussian as
reionization progresses~\citep{WM07,harker2009}, particularly on small scales when HII regions are formed~\citep{furl2004b,bharadwaj2005}. As a result, the PS does not provide a complete statistical description of the reionization process.
Some additional
information could be captured by skewness in the probability distribution~\citep{WM07,harker2009}. Estimates in these papers indicate that MWA and LOFAR will have sufficient
sensitivity to detect the evolution in the skewness, though foreground subtraction has yet to be considered in depth.

 \citet{ichikawa2009} extended these ideas by
calculating the evolution of the probability distribution of intensity
fluctuations using a 
numerical simulation of cosmic reionization. 
The intensity
distribution has a Gaussian shape at high redshift, and then develops
an exponential tail at low intensity as HII regions form. Towards the
end of reionization as HII regions dominate the volume, the
distribution becomes strongly peaked at a neutral fraction of zero. 
%

\subsection{Cosmology with EoR observations}
\label{EoRCosmo}

While the prime motivation for EoR experiments has been measurement
of the PS to probe the astrophysics of the reionizing
sources, a number of authors have explored the
possibility of measuring cosmological parameters from the shape of the
underlying matter PS. 

\citet{bowman2007} considered the possible
constraints on cosmological parameters that would be available from
measurements of the spherically averaged 21~cm PS. They
found that even in a scenario where the PS was dominated
by density fluctuations (i.e. before reionization is underway), planned 
telescopes like the MWA would not be able to make
interesting cosmological measurements. Similarly pessimistic results
were presented by \cite{santos2006} who found significant degeneracies
between astrophysical and cosmological parameters in the angular
PS. Moreover, during the the EoR, the
PS of 21~cm brightness fluctuations is shaped mainly by the
topology of ionized regions, rather than by the PS of
matter density fluctuations which is the quantity of cosmological
interest~\citep{McQ+06, S+07, iliev2007}. With respect to measurement
of the mass PS, the appearance of HII regions complicates the interpretation of a measured PS by generating
scale dependence in the effective bias that relates the 21~cm
PS to the linear mass PS.

On the other hand, peculiar velocities of the 21~cm emitting gas are sensitive only to the mass density and not to the ionization structure. As a result, the line-of-sight anisotropy of the 21~cm PS due to peculiar velocities may be used to separate
measurements of the density PS from the unknown details of
the astrophysics driving the growth of HII regions \citep{Barkana:2005p1804}. 
Although the dependence of the PS on
angle is nontrivial (see equation~\ref{P21}), the cosmological signal could be
extracted by isolating the term proportional to $\mu^4$. However \citet{McQ+06} 
found that planned low frequency arrays like MWA and LOFAR will not be
sensitive to the $\mu^4$ term, and so could only provide useful
cosmological constraints in a situation where density
fluctuations dominate the 21~cm PS.

More recently, \citet{mao2008} have shown that parametrization of the
ionization PS could be used to model the full 21~cm
PS in order to extract cosmological constraints. The
required parametrization is motivated by the generic form expected for
the ionisation PS,
and in practice adds several free parameters to the cosmological
fit. Although the method relies heavily on the applicability of the reionization model, these authors show that marginalizing over the additional
parameters is more effective than using only the $\mu^4$ term in the 21~cm PS (equation~\ref{P21}). Indeed, \citet{mao2008} show that parameterization of the ionizations structure may provide constraints on the cosmology that are almost as tight as a case where 21~cm
tomography measures the matter PS directly. However
cosmological constraints that are competitive with existing
measurements (from CMB etc.) will still require larger arrays than those currently planned.

\subsection{The 21~cm fluctuations from the post reionization IGM}
\label{aftereor}

Until recently, the conventional wisdom has been that the 21~cm signal
disappears after the end of the EoR, because there is
little HI left through most of intergalactic
space. However the redshifted 21~cm
emission from a volume of IGM is sensitive to $\bar{x}_{\rm HI}$, which is non zero at all cosmic epochs (e.g. Figure~\ref{fig_WL}). The following simple estimate can be used to
demonstrate that 21~cm emission should be significant even after
the EoR is completed. 

Observations of DLAs out
to a redshift of $z\sim4$ show the cosmological density parameter of
HI to be $\Omega_{\rm HI}\sim10^{-3}$ \citep{prochaska2005}. This 
corresponds to a mass-averaged HI fraction of $\bar{x}_{\rm HI}\sim0.03$, 
%
and a brightness temperature contrast of redshifted 21~cm
emission of $\Delta T\sim0.5$mK (see equation~\ref{Tb}). Moreover on $\sim10$
co-moving Mpc scales the typical amplitude of density
fluctuations at $z\sim4$ is $\sigma\sim0.2$. Hence, we expect
fluctuations in the radiation field of $\sim0.1$mK after reionization on scales relevant for upcoming 21~cm experiments.
As seen in Figure~\ref{fig_WL}, these fluctuations are only an order of magnitude or so smaller than
the largest fluctuations predicted at any time during the entire EoR. At the same time the sky brightness (which limits the observational sensitivity, see \S~\ref{astroForegrounds}) will be $\sim5$ times smaller owing to the frequency dependence of the foreground synchrotron emission.  Thus,
detectability of fluctuations in 21~cm emission may not decline
substantially following the overlap epoch. The post-reionization HI could therefore be studied via 21~cm intensity mapping using telescopes like those described in \S~\ref{ct}.
%

The detectability of the 21~cm PS after reionization was discussed by \citet{khandai2009}. These authors used a numerical simulation to predict the statistical signal of 21~cm fluctuations in the post-reionization IGM, and estimated its detectability by telescopes like the MWA and GMRT. \citet{khandai2009} find that a combination of these arrays offer good prospects for detecting the 21~cm PS over a range of redshifts in the post reionization era.
Importantly, a statistical detection of 21~cm fluctuations due to discrete,
unresolved clumps of neutral gas has already been made \citep{pen2009}
through cross-correlation of the HIPASS \citep{barnes2001} 21~cm
observations of the local universe with galaxies in the 6 degree field
galaxy redshift survey \citep{jones2005}. This detection represents an
important step towards using 21~cm surface brightness fluctuations to
probe the neutral gas distribution in the IGM (both during and after
reionization).

%

\subsubsection{The 21~cm power spectrum after reionization}

The form of the post-reionization
PS can be derived in analogy with the PS during
the EoR \citep{wyithe2009b}. To begin, the fluctuation in ionization fraction may be
written as
\begin{equation}
\label{dx}
\delta_{x}\approx[{\bar{x}_{\rm HI}}/{(\bar{x}_{\rm HI}-1)}]\,\left[b(1-K(k))-1\right]\delta.
\end{equation}
%
%
where $b$ is the clustering bias of the HI relative to the underlying density field. 
The function $K$ is introduced
to quantify the effect that the ionizing background should have on the
neutral fraction. 
Fluctuations in the HI content of individual DLAs can be shown
to be proportional to fluctuations in the ionizing background (which has a small effect on the density at which the DLA becomes self shielded).
By convolving the real space density field with a filter function it
is then possible to calculate the function $K$, which
becomes
\begin{equation}
K(k)=K_o \frac{\arctan(k\lambda_{\rm mfp})}{k\lambda_{\rm mfp}},
\end{equation}
where $k=|\vec{k}|$, and $\lambda_{\rm mfp}$ is the ionizing photon mean-free-path. In \citet{wyithe2009b} it was shown that for
realistic DLA profiles the constant $K_o$ has a value smaller than
$K_o\sim0.01$. 

From equation~(\ref{dx}) we see that the fluctuation in ionization is proportional to the fluctuation in density up to a weak (and simple) scale dependence, yielding an expression for the direction dependent post-reionization 21~cm PS
\begin{equation}
\label{DLAPS1}
P_{\Delta T}(b)=\mathcal{T}_{\rm b}^2\bar{x}_{\rm HI}^2\left[b(1-K(k))+\f\mu^2\right]^2P_{\delta\delta},
\end{equation}
where terms are defined as per equation~(\ref{P21}).
Thus the 21~cm PS after reionization has a form that is very similar to the familiar galaxy PS in the regime of scale independent halo bias.

\subsubsection{Cosmology with 21~cm fluctuations after the EoR}

The simplicity of equation~(\ref{DLAPS1}) allows near direct
observation of the mass-PS, and hence the full 21~cm
PS to be utilized in deriving cosmological
constraints. This compares favorably with the EoR,
during which either the $\mu^4$ term alone can be utilized
~\citep{Barkana:2005p1804}, or a functional form for the ionization
PS must be assumed ~\citep{mao2008}. The
resulting utility of the 21~cm PS from HI in the post-reionization Universe has
been pointed out by a number of
authors.

For example, measurement of the scale of baryonic acoustic
oscillations using 21~cm fluctuations in the post-reionization Universe
was discussed by \citet{WLG08} who focused on redshifts $z>3$ where
telescopes of the design of MWA could operate, and by
\citet{chang2008} who looked at measurement of the oscillations in the
redshift range $z<2$ where measurements best constrain conventional
dark energy. \citet{chang2008} argued that a radio telescope with a
diameter of $\sim200$ wavelengths would provide an efficient method for
surveying huge volumes, and so provide constraints on dark
energy parameters that would be competitive with planned galaxy redshift
surveys.

The studies of \citet{loeb2008} and \citet{visbal2008} focused on
cosmological constraints from studies of the full PS using
futuristic arrays beyond the first generation. \citet{loeb2008} noted
that a dedicated HI cosmology observatory could probe a number of independent
modes two orders of magnitude larger than currently available, and that
this would enable a cosmic-variance limited detection of the signature
of a neutrino mass as small as 0.05 eV. \citet{visbal2008} 
provide a
detailed forecast of the constraints on cosmological parameters that
would be achievable, and find that HI cosmology telescopes
dedicated to post-reionization redshifts may yield significantly
better constraints than next generation Cosmic Microwave Background
(CMB) experiments. In a complementary study, \citet{bharadwaj2009}
focused on the possibility of using 21~cm fluctuations to determine the
redshift-space distortion parameter, as well line of sight verses
transverse co-ordinate distances. Using the Alcock-Paczynski effect
combined with sensitivity estimates corresponding to forthcoming low-frequency
arrays, these authors predict constraints on dark energy parameters that would be at a
precision comparable to state of the art type-Ia supernova observations and
galaxy redshift surveys, but across a wide range in redshift.

\subsubsection{Galaxy evolution with 21~cm fluctuations after the EoR}

In addition to cosmology, intensity mapping instruments will also provide an important new probe for studying the global properties of the galaxy population. For example, \citet{wyithe2008b} shows that measurement of the PS of 21~cm intensity fluctuations (on linear scales) 
could be used to measure both the galaxy bias for DLA host galaxies,
and the cosmological HI
mass density. On smaller scales, simulations suggest that the HI density field is highly biased so that individual rare peaks might be detected, facilitating a direct estimate of the cosmic HI mass density~\citep{bagla2009}. The evolution of the cosmic HI mass density and its environmental dependence will provide crucial evidence in our developing understanding of the role of neutral gas in galaxy formation.

\begin{figure}
\begin{center}
\includegraphics[width = 5 in]{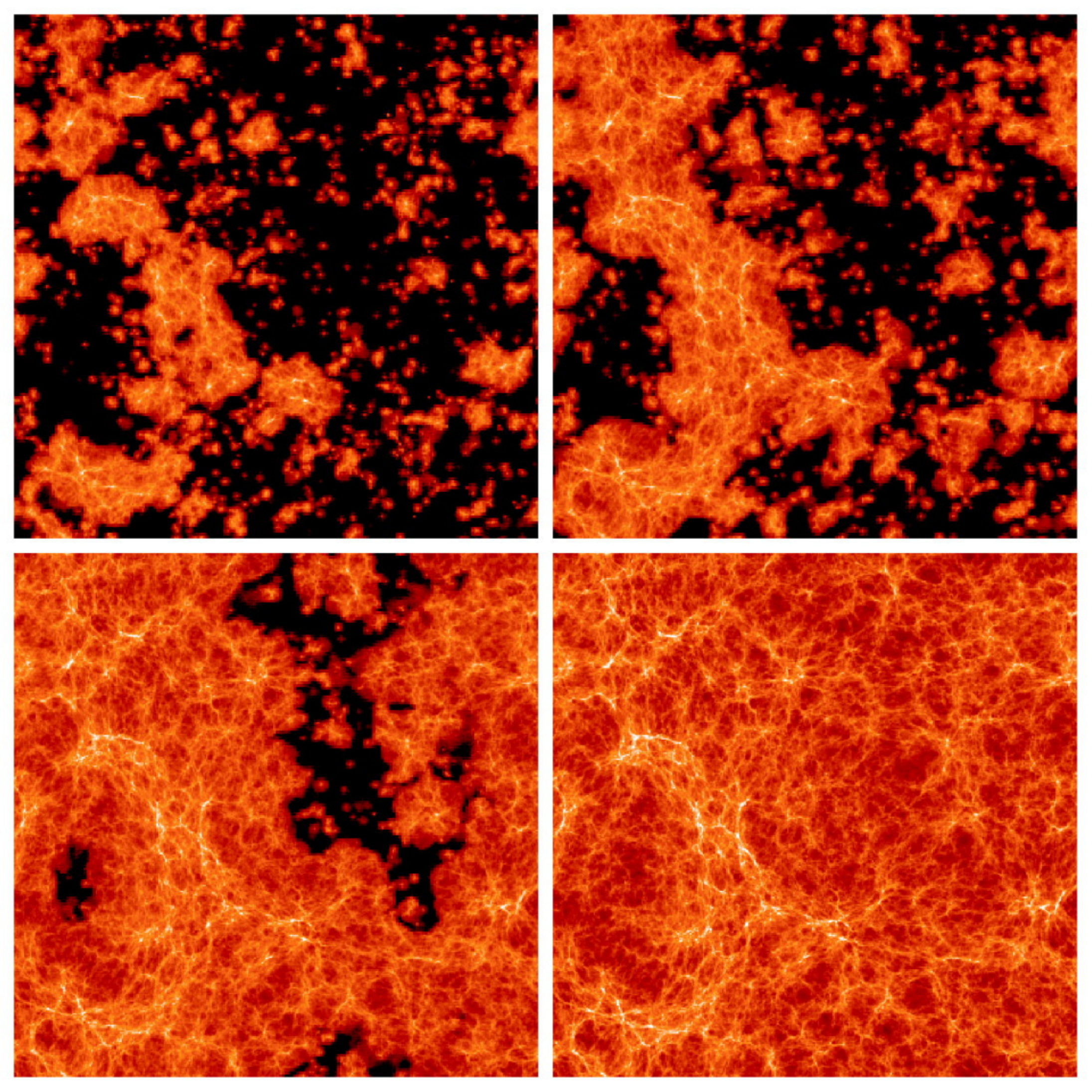}
\caption{ From \citet{trac2007}. Image showing the HII density (brighter indicates higher density of ionized hydrogen) in 1 Mpc $h^{-1}$ deep slices through a simulation box (50$h^{-1}$ Mpc on a side) at redshifts $z$ = 9 (top left), 8 (top right), 7 (bottom left), and 6 (bottom right). The volume-weighted mean H I fractions are $\langle F_{\rm HI}\rangle_{V} = 0.67$, 0.46, 0.12, and $1.5\times10^{-4}$, respectively. At high redshifts the neutral IGM appears as large dark regions with bubbles of neutral hydrogen around the first sources. These bubbles eventually percolate, reionizing the universe by the last panel. In the last panel the cosmic web can be seen as a tracer of higher density gas. Neutral hydrogen survives in dense galactic systems, enabling the detection of HI fluctuations after reionization. Note that the image is plotted in terms of HII density which accentuates the appearance of HII regions since the fluctuations of HI in the neutral IGM (which are in fact the 21~cm observable) are not displayed. }
\label{Trac1}
\end{center}
\end{figure}

\begin{figure}
\begin{center}
\includegraphics[width = 5.25 in]{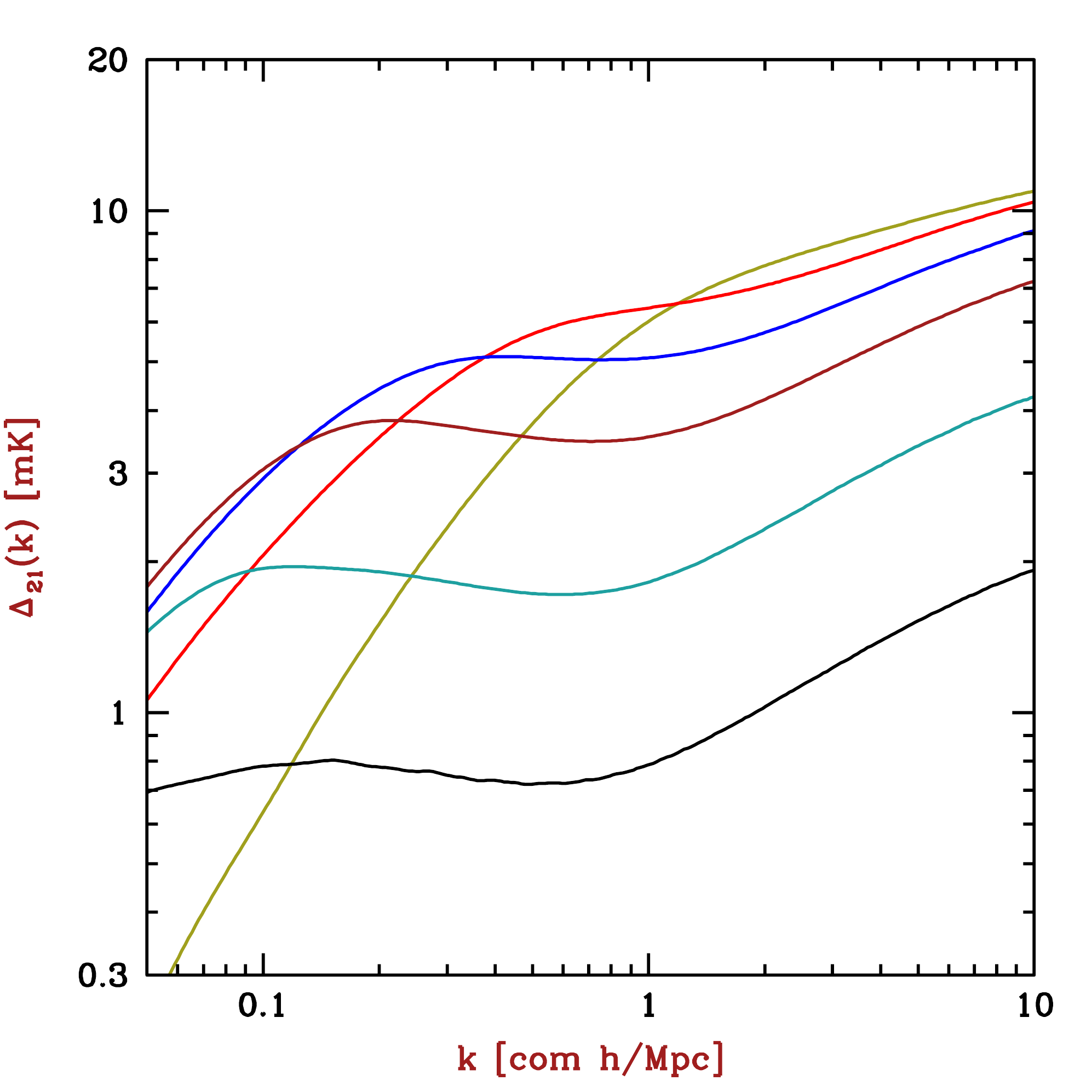}
\caption{ From \citet{barkana2008}. Evolution of dimensionless 21~cm PS $\Delta_{21}\equiv \sqrt{k^3 P(k)/(2\pi^2)}$ throughout the EoR,
for a model that sets $\bar{x}_i$ = 98\% at $z = 6.5$ with a minimum
circular velocity for ionizing sources of $V_{\rm c} = 35$ km/s. The models shown have $\bar{x}_i$ = 10\%, 30\%, 50\%, 70\%, 90\%, and
98\% (from top to bottom at large $k$) and show the characteristic PS shoulder from HII regions moving from small to large scales as reionization progresses. }
\label{Barkana2008Fig4}
\end{center}
\end{figure}

\begin{figure}
\begin{center}
\includegraphics[width = 5.251 in]{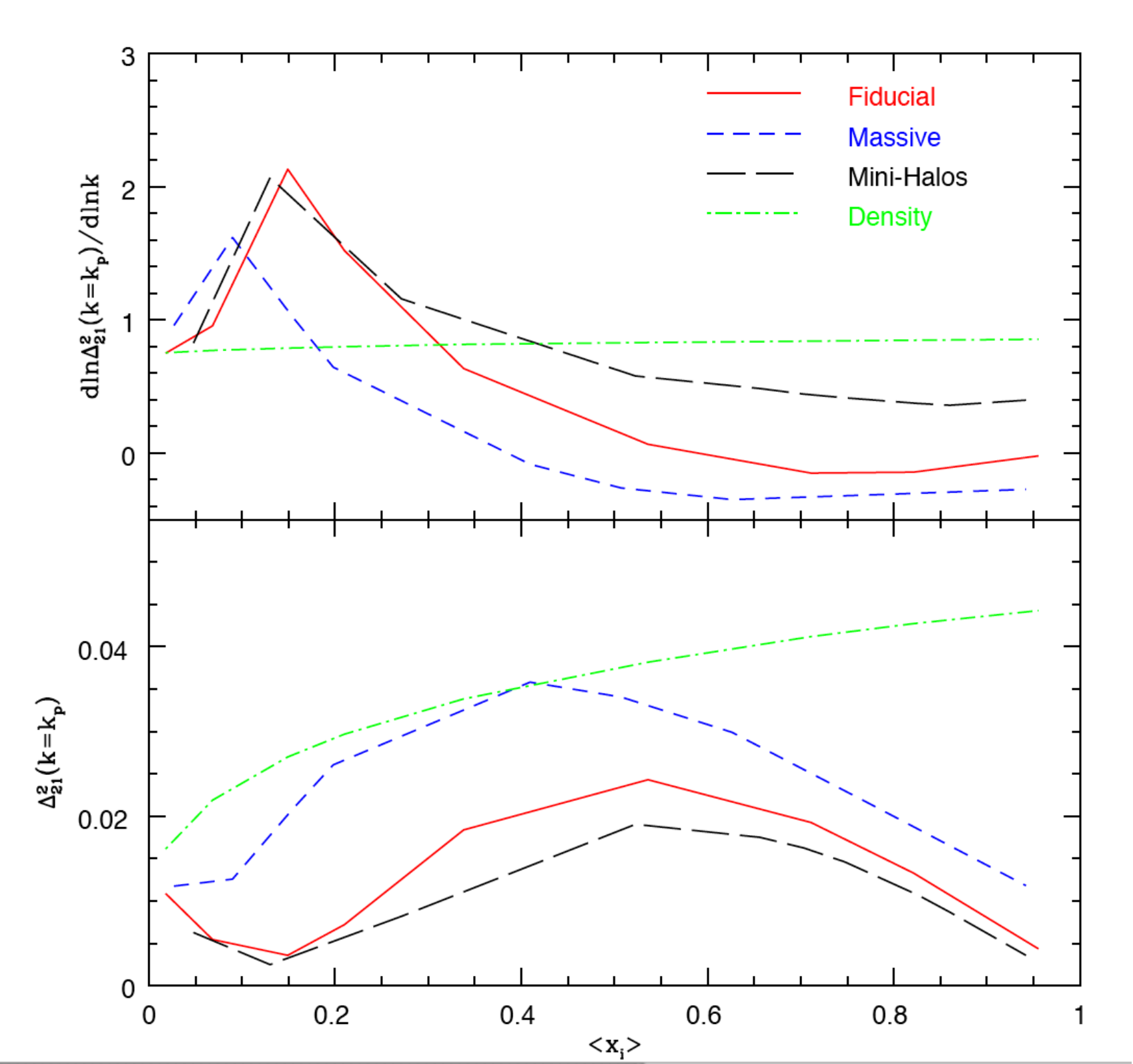}
\caption{From \citet{lidz2008}. Amplitude and slope of model 21~cm power spectra
$\Delta_{21}^2\equiv k^3 P(k)/(2\pi^2)$ as a function of ionization fraction (in units of $(28 [(1 + z )/10] mK)^2$. Bottom: Amplitude of the 21
cm PS, at the pivot wavenumber ($k = 0.4$h Mpc$^{-1}$)
for MWA observations in a fiducial (solid red line), rare source
(blue short-dashed line), and mini-halo models (black long-dashed
line), plotted as a function of ionization fraction \citep[see][for further details]{lidz2008}. For comparison, the green dot-dashed
line shows the amplitude of the density power
spectrum obtained by mapping redshift to ionization fraction. Top: Slope of the 21~cm PS at
the pivot wavenumber for our three models, as well as the density
PS slope. }
\label{lidzfig2}
\end{center}
\end{figure}

\begin{figure}
\begin{center}
\includegraphics[width = 5.25 in]{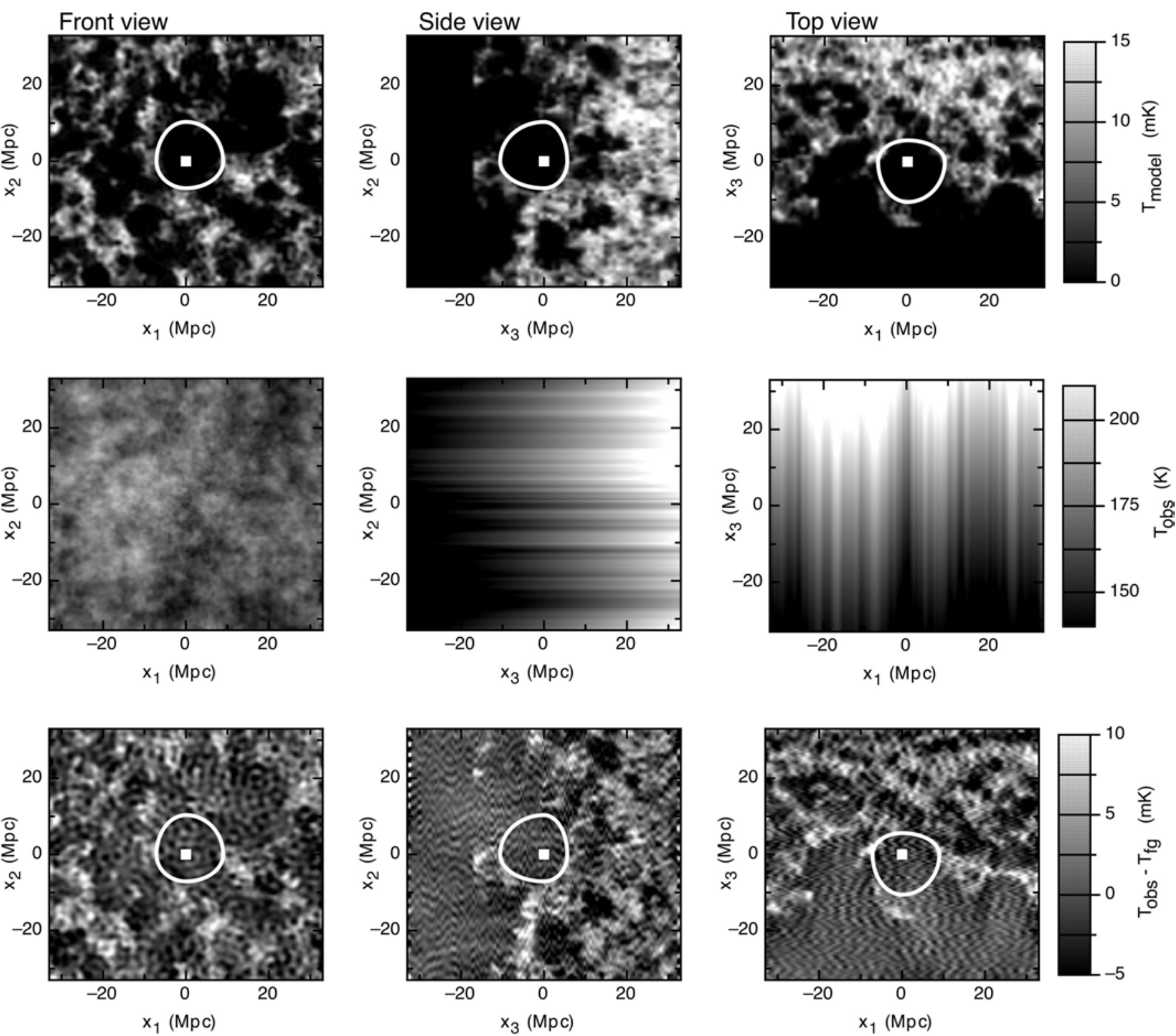}
\caption{From \citet{GWPO08}. A quasar HII region in an evolving IGM. Three aspects of the HII region are shown, with slices through the center of the box when viewed from the front, top and side. The quasar was assumed to contribute ionization equivalent to an HII region of radius $R_{\rm q,tot}=34$ co-moving Mpc, and to be centered on $z=6.65$, which is also the redshift at the center of the simulation box. Each slice is $6\,$Mpc thick, which corresponds to $\sim3\,$MHz along the $x_3$-axis. In observed units, the cube is $\sim3.3$ degrees on a side and $33\,$MHz deep. The model, foreground plus model, and observed maps following foreground removal are shown in the upper, central and lower panels respectively. The shape of the HII region extracted from the foreground removed cubes is also plotted~\citep[see][for details]{GWPO08}. The mass-averaged IGM neutral fraction was assumed to be $\bar{x}_{\rm HI}=0.15$. }
\label{Wyithe1}
\end{center}
\end{figure}

\def\r{\mathbf {r}}
\def\k{\mathbf{k}}
\def\u{\mathbf{u}}

\section{Observations}
\label{ObsSec}

The last few years have seen major advances in instrumentation for observing HI fluctuations. The major existing low frequency observatory has been significantly upgraded to search for the EoR power spectrum, and there are three new EoR observatories currently under construction. In addition, a number of HI intensity mapping machines are in the design and prototype phases, and we expect at least one of these to start construction in the next few years. These HI machines are pushing the state-of-the-art in widefield observations, digital data handling and correlation, precision instrumental and atmospheric calibration, and high fidelity foreground subtraction. In this section we review the current state of the field with an emphasis on the astrophysical and observational effects that drive instrument design, and the solutions different groups have developed. We start with a brief introduction to interferometry (\S \ref{InterferometryIntro}) and the sensitivity of interferometers to the HI power spectrum (\S \ref{HIsensitivity}). We then discuss instrumental calibration in \S \ref{calibration} and end with foreground subtraction in \S \ref{foregrounds}.

\subsection{Introduction to interferometric measurements}
\label{InterferometryIntro}

\begin{sidewaysfigure*}
\includegraphics[width=7.5in]{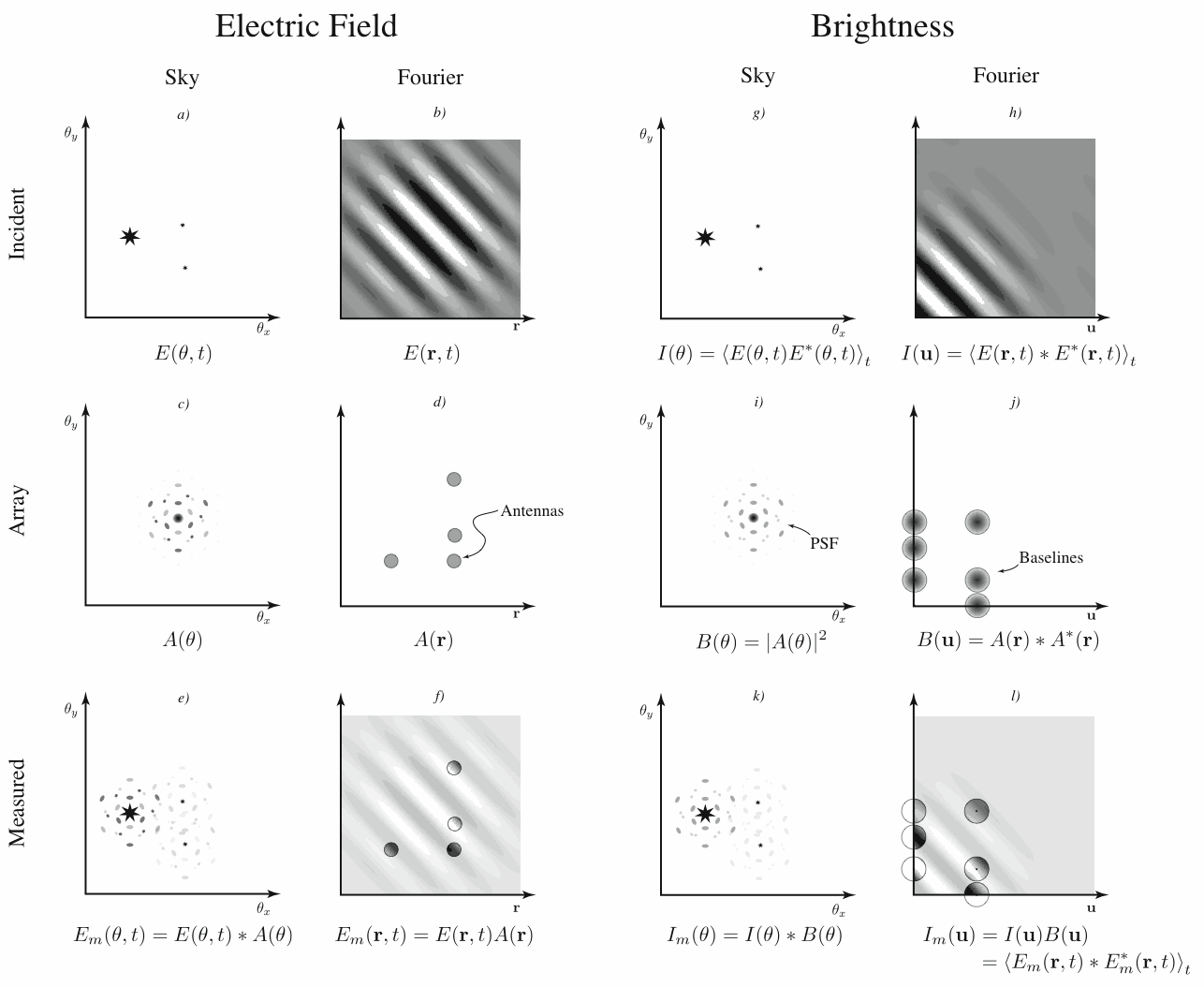}
\caption{\scriptsize{This figure pictorially depicts the fundamental measurement process of interferometry. The six lefthand panels ($a$-$f$) describe the electric field while the right six panels correspond to the same operations in terms of brightness. Within each set of six panels the lefthand column is expressed in sky coordinates ($\sky$) while the righthand column shows the Fourier coordinate representation of the same relationship. The rows of the figure correspond (top to bottom) to the incident or true radiation pattern, the properties of the interferometric array, and the resulting electric field or brightness measurements. The number of antennas and the point spread function have been simplified for illustration purposes.}}
\label{EIsquares}
\end{sidewaysfigure*}

All of the proposed HI measurements of the EoR and the dark energy equation of state utilize radio interferometers, and interferometers are best understood in the Fourier domain. 
In an effort to build a conceptual understanding of interferometers, it is instructive to start with the electric field as seen by the array and build up to the final brightness measurement.  We will limit our discussion to narrow fields-of-view and antennas in a plane to focus on the key relationships, and the text explanation is paralleled by a pictorial description in Figure \ref{EIsquares}.  The interested reader is encouraged to look at \citet{SynthImagII}, \citet{Rau:2009p4084} and \cite{Carozzi:2009p3904} for complete mathematical descriptions of interferometry, including curved sky, widefield, and non-planar array effects.

The direction dependent electric field at a reference position $E(\sky,t)$ can be integrated to form the total electric field at the reference position $E(t)$ (small FoV limit), or propagated to nearby locations (in the far-field limit) using
\beq
E(\mathbf {r}, \sky, t) = E(\sky, t)e^{2\pi i\ \sky \cdot \mathbf {r}/\lambda}.
\label{Eprop}
\eeq
Integrating over the sky to determine the total electric field at each location, we obtain the fundamental relationship of interferometry:
\beq
E(\r,t) = \int E(\sky, t)e^{2\pi i\ \sky \cdot \r/\lambda}\ d^{2}\sky.
\label{IntEq1}
\eeq
The electric field as a function of position on the ground is the Fourier transform of the electric field coming from the sky (see panels $a$ and $b$ of Figure \ref{EIsquares}). This happy coincidence between the electric field propagator and the Fourier transform forms the basis of interferometry.

While radio antennas directly measure the electric field, what we are really interested in is the sky brightness $I(\sky)$. Even over very narrow ranges of frequency, the phase of the incident electric field will change randomly (not a delta-function in frequency space). However, Equation \ref{Eprop} still holds and the electric field at two locations will be delayed by a constant amount. The sky brightness is the time-average square of the electric field:
\beq
I(\sky) = \left \langle E(\sky,t)E^{*}(\sky,t) \right \rangle_{t} = \left \langle \left | E(\sky,t) \right |^{2} \right \rangle_{t}.
\label{Iskydef}
\eeq
Using the relation that a multiplication in one Fourier space is a convolution in the dual space ($HG = h * g$), we can express the sky brightness in Fourier coordinates either by taking the auto-convolution of $E(\r)$ in analogy with equation \ref{Iskydef}, or by directly taking the Fourier transform of $I(\sky)$. This gives us
\beq
I(\u) = \left \langle E(\r,t)*E^{*}(\r',t) \right \rangle_{t},
\eeq
where $\u = \Delta \r$ and this is shown pictorially in  panels $g$ and $h$ of Figure \ref{EIsquares}.

The middle row of Figure \ref{EIsquares} describes the measurement process. We can describe our array as the locations on the ground $\r$ where the electric field is collected---for an array of dishes this looks like a series of small circles at the location of each antenna as in panel $d$. The inverse Fourier transform of the antenna distribution $A(\r)$ gives us the electric field point spread response $A(\sky)$, the square of which gives us the usual brightness point spread function $B(\sky)$. The point spread function (PSF) is equivalent to the power response the array would measure for a single point source. We can form the last panel in the row ($j$) either by Fourier transforming the point spread function, or more conceptually by taking the auto-convolution of the antenna distribution (panel $d$). Due to the auto-convolution there is a brightness measurement for every \emph{pair} of antennas ($N(N-1)/2$ pairs). Furthermore the size of each measurement region is $\sim$twice the diameter of the antennas shown in $A(\r)$ (for clear apertures $A(\r)$ is a circular tophat at each antenna, and $B(\u)$ is a circular cone at the location of every antenna pair/baseline).

The last row of Figure \ref{EIsquares} then represents the observations. In the Fourier space representations the array samples a subset of the incident electric field/brightness as shown in panels $f$ and $l$. Since this sampling is a multiplication in the Fourier domain, it is equivalent to convolving the sky (panels $a$ and $g$) with the electric field or brightness point spread functions (panels $c$ and $i$).

While most astronomers are comfortable with convolving the input sky with a point spread function, working in the Fourier space is less familiar. However, an interferometer works by measuring the electric field at each antenna, which is then digitized and converted into cross-power \emph{visibilities} in the correlator. 
Each of the visibilities corresponds to the brightness measurement
\beq
I_{m}(\u) = \left \langle E_{m}(\r,t)*E^{*}_{m}(\r,t) \right \rangle_{t}
\label{panelL}
\eeq
at one of the antenna pairs, as shown in panel $l$ of Figure \ref{EIsquares}. Each of the visibilities measured by the array is equivalent to measuring the underlying Fourier mode $\u$. An instrument designer has a lot of flexibility with an interferometer, because they can tune the Fourier modes that are measured by changing the arrangement and separation of the antennas. For cosmology, this best examples of this are the DASI \citep{Leitch:2002p4411} and CBI experiments where the designers maximized their signal-to-noise by choosing antenna distributions that only measured the angular Fourier modes of interest. In the next section we will see how changing the antenna configuration affects the sensitivity of arrays to cosmological HI signals.

In practice the optics of the antenna sum the measured electric field $E_{m}(t, \r)$ and squeeze it into a single RF cable $E'_{i}(t)$ (integral over the disk of an antenna in panel $d$). The electric field is split into many narrow frequency channels in the first stage of the correlator, then the cross-power visibilities between antenna pairs are calculated using
\beq
v_{ij}  = \left \langle E'_{i}(t)E'^{*}_{j}(t) \right \rangle_{t}
\label{correlatorVis}
\eeq
in each narrow frequency channel.\footnote{The procedure described is the FX correlator algorithm (frequency then multiply), and is used for all current EoR instruments. There are alternate but equivalent algorithms such as the XF algorithm used in the VLA correlator.}
The measured visibility of a pair of antennas $v_{ij}$ is equivalent to the average over position ($\u$) and the frequency channel ($\Delta f$)
\beq
v_{ij}  = \left (\frac{1}{\Delta f A_{e}}\right )\int_{\Delta f}\int I(\u)B_{ij}(\u)\ d^{2}\u\ df
\label{visSens}
\eeq
for the corresponding baseline in panel $l$, and we have used the relation $\int f(x)*g(x) dx = \left ( \int f(x)dx \right )\left ( \int g(x)dx \right)$ to convert the integral over $\left \langle E_{m}(\r,t)*E^{*}_{m}(\r,t) \right \rangle_{t}$ in Equation \ref{panelL} into the multiplication in Equation \ref{correlatorVis}. The visibility cross-power has units of mK or Jy, depending on the units chosen to express the sky brightness $I(\sky)$, and the uncertainty in the measurement is given by
\beq
v_{\rm rms} (\u) = \frac{\lambda^{2}\Tsys}{A_{e}\sqrt{\Delta f \tau}},
\label{visRMS}
\eeq
where $\tau$ is the integration time for that spatial mode, $\Tsys$ is the system temperature (equivalent power of a resistor on the cable at temperature $\Tsys$), $A_{e}$ is the effective area of an antenna, and $\lambda$ is the observing frequency. \footnote{$\Tsys$ is the dual polarization intensity system temperature in this context. If working in Jy units multiply by $2k_{\rm B}/\lambda^{2}$. The the constants in Equations \ref{visSens} and \ref{visRMS} implicitly assume  $\int B(\sky) = A_{e}$ \citep{Rohlfs,Morales:2004p803}.}

\begin{figure}
\begin{center}
\includegraphics[width=5.25in]{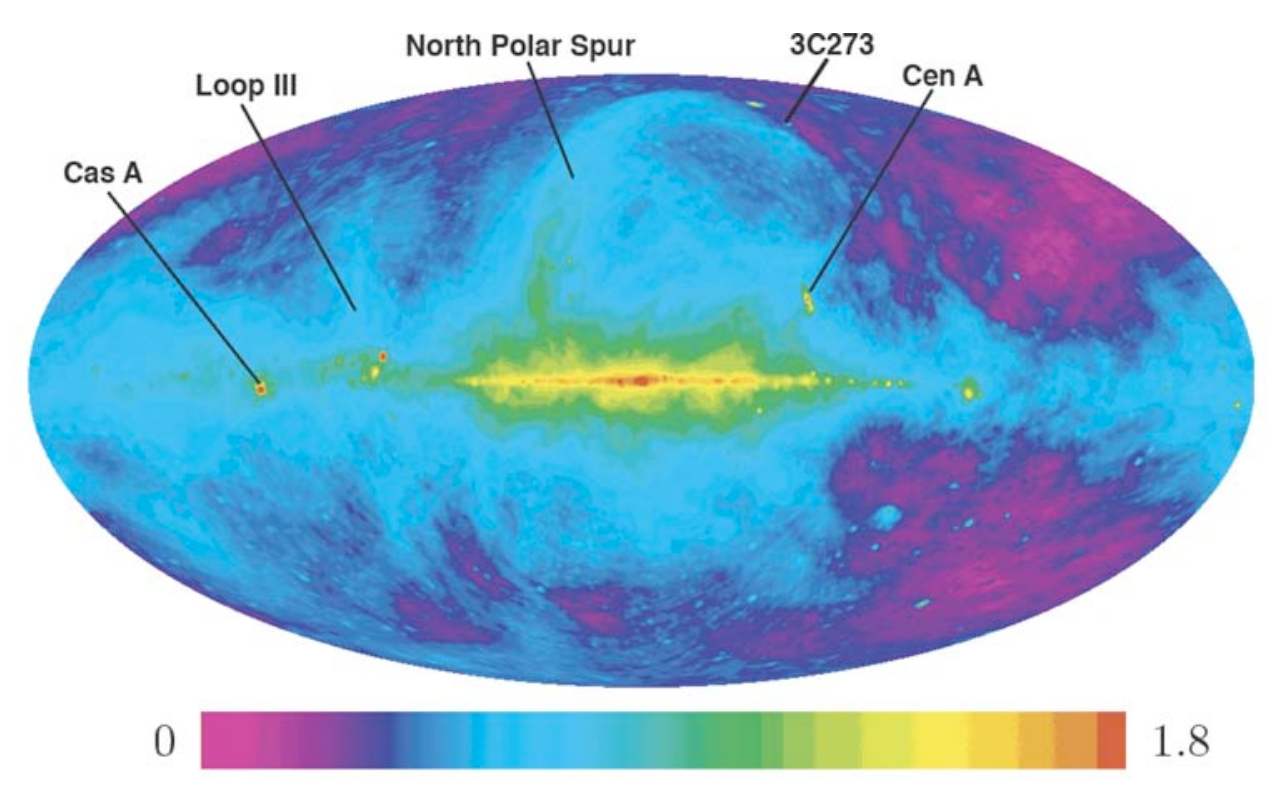}
\caption{A plot of the synchrotron emission from \citet{deOliveiraCosta:2008p3515}, on a logarithmic relative amplitude scale (amplitude depends on frequency). The brightness of the of the galactic plane is easily seen, as are large scale features such as the North Polar Spur and extremely bright extra-galactic sources such as Cassiopeia A. Oliveira-Costa's model can be scaled to any of the frequencies of HI cosmology and includes dust emission and spectral curvature effects.}
\label{AngSynch}
\end{center}
\end{figure}

The system temperature is the sum of the sky brightness---primarily galactic synchrotron at EoR and post-reionization frequencies---and the analog receiver noise ($\Tsys = T_{\rm sky} +T_{\rm rec}$). The intensity of the synchrotron emission scales with frequency as approximately $f^{-2.6}$, and varies by orders of magnitude from the central regions of the galaxy to high latitudes \citep[see Figure \ref{AngSynch} from][]{deOliveiraCosta:2008p3515}. While good uncooled low noise amplifiers can have noise temperatures as low as 15~K, the full receiver noise typically ranges from 40~K to 100~K depending on the impedance match of the antenna and how much of the thermal glow from the ground leaks in. For EoR experiments it is common to have fairly high receiver temperatures to keep the antenna costs down, as the system temperature is dominated by the several hundred degree sky contribution, whereas 
intensity mapping instruments operating at lower redshift
must pay much closer attention to the analog systems as the sky temperature is only 5--10~K.

The last step in accurately describing the sensitivity of an interferometer is to include sky rotation. While a single baseline measures one Fourier mode $\u$ instantaneously, the apparent location of the baseline moves as the source tracks overhead (in Figure \ref{EIsquares} panel $l$ the baseline locations change). Conceptually this is easiest to see by envisioning the array painted on the surface of the earth, as viewed by the celestial source.  From the perspective of the celestial source the orientation of the array appears to rotate and foreshorten as the Earth spins (if neither the array or celestial source are at the equators). Over the course of a multi-hour observation the movement of the baselines through $\u$ means that a many more Fourier modes are observed. As the sampling of Fourier plane improves the quality of the recovered image increases dramatically, an effect called `rotation synthesis.'


However, the uncertainty on a single Fourier mode is related to how long that particular mode was observed, not the total observation time. To determine the signal-to-noise per mode, we can sample the $\u$ plane on a regular grid and add the measured correlations to the appropriate sample location (this is done with a convolving kernel and is referred to as `gridding' in the interferometry literature). The noise for a single sample $v(\u)$ is the total weighted integration time---how long that spatial mode was observed. Alternatively we can pixelate the $\u$ plane on the scale of the instrumental beam $B(\u)$ and count the amount of time baselines spent within that $\u$ pixel.

There are a number of important effects which have been ignored in this simplified introduction to interferometric observations, including multi-frequency synthesis, non-coplanar antenna locations (3D $\u$), and curved sky effects. The interested reader is referred to the review by \citet{Rau:2009p4084}, the VLA summer school proceedings by \citet{SynthImagII}, and the papers by \citet{Bhatnagar:2008p3407,Cornwell:2008p3151} and \citet{Morales:2009p3737}.

\subsection{Sensitivity of an HI interferometer}
\label{HIsensitivity}


The HI signals from both the Epoch of Reionization and lower redshifts are inherently three dimensional, with the observed frequency directly providing the emission redshift. The emission from the cosmological volume ($\{x,y,z\}$ in cMpc) is mapped by the redshift and angular diameter distance into angular position and frequency $\{\sky_{x}, \sky_{y}, f\}$ \citep{Hogg:1999p923}. However, as discussed in the previous section interferometric measurements are performed in Fourier space. A typical radio interferometer thus measures the Fourier modes $\{\u_{x}, \u_{y}\}$ in several thousand frequency channels ($\Delta f \approx 10$~kHz), and can be pictured as simultaneously measuring panel $l$ of Figure \ref{EIsquares} in several thousand closely spaced frequency planes. In many ways, HI measurements are very strange as the $\{\u_{x}, \u_{y}, f\}$ measurements can be interpreted as cosmological wavenumbers in the angular direction (inverse distance) and line-of-sight distance in the frequency direction $\{k_{x}, k_{y}, z\}$ with units of $\{{\rm cMpc}^{-1},{\rm cMpc}^{-1},{\rm cMpc}\}$. The visibility measurements can either be Fourier transformed in the angular directions to form an image volume (with PSF artifacts), or can be Fourier transformed in the line-of-sight (frequency) direction to form a three dimensional wavenumber cube $\{k_{x}, k_{y}, k_{z}\}$ \citep{Morales:2004p803}. 

The three dimensional $k$-space is particularly useful for power spectrum measurements. In the absence of cosmic evolution and velocity distortions, the power spectrum signal is spherically symmetric in the three dimensional $k$-space $\{k_{x}, k_{y}, k_{z}\}$. This approximate spherical symmetry is due to the isotropy (rotational invariance) of space, and its importance for HI cosmology measurements is difficult to overstate. Measurements of the one dimensional power spectrum are performed by measuring the variance (average of squares) of all the measurements within a spherical shell in $k$-space ($|k|  \pm \Delta k/2$); astrophysical foregrounds have a different shape in $k$-space which enables their removal (\S \ref{astroForegrounds} and \S \ref{obsForegrounds}); and the density of measurements in $k$-space drives the design of HI cosmology arrays (discussed later this section). This is the same rotational symmetry which leads to averaging over angular $m$ modes in CMB measurements, just translated from the two-dimensional surface of the CMB to the three dimensional volume of HI cosmology measurements.

Of course the spherical symmetry is only approximate---the universe evolves very quickly during the EoR, there is large scale gravitational infall \citep{Barkana:2005p1804}, and non-linear velocity effects (`fingers of God') become significant at the lower redshifts targeted by HI intensity mappers. For HI intensity mapping this is an advantage. Because the cosmological models predict the apparent angular diameter, the relative line-of-site distance ($\Delta z$), and the velocity effects, these $k$-space asymmetries can be used to tighten the cosmological constraints \citep{visbal2008}. Unfortunately the theoretical EoR models are significantly more complex, and the evolution of the ionized bubbles and the parameters which control this evolution are much less certain (see \S \ref{signal}). This uncertainty leaves the EoR observer without a definite model to test---there are no widely agreed upon parameters which we can use in a Fisher matrix analysis. To date the response has been to take a narrow range of redshift $\Delta z \approx 0.5$ (about 8~MHz), and assume that cosmic evolution is relatively minor over this interval. Isotropy is then used to generate a spherically averaged one dimensional power spectrum within each $\Delta z \approx 0.5$, resulting in a coarse power spectrum `movie' showing the dynamics of reionization and how the bubbles expand and overlap (see Figures \ref{Barkana2008Fig4} and \ref{lidzfig2}). The advantage of this approach is that it is largely model independent, and allows the parameters of a wide variety of theories to be constrained. Once the theory matures, the EoR will become more like the CMB with specific parameters that we can constrain. 

For power spectrum measurements, the signal $P(k)$ is independent for each Fourier mode in the three dimensional $k$-space.\footnote{There are non-Gaussian and non-stationary statistical correlations during the EoR imprinted by the bubbles, but no one has figured out how to measure these statistical properties/topology with first generation arrays.} This statistical independence makes $k$-space the natural frame for calculating the sensitivity of an interferometer. (Here we are closely following the papers by \citet{Bowman:2006p163}  and \citet{mcquinn2007}.) In the natural $\{\u_{x}, \u_{y}, f\}$ space of an interferometric measurement, the uncertainty within a given `pixel' of $\u$ or $\kperp$ space is given by Equation \ref{visRMS}, where the size of the pixel is given by the width of the antenna $\times$ 2 ($B(\u) = A(\r)*A^{*}(\r)$), the depth of the pixel is given by the frequency resolution ($\Delta f$ or $dz$), and the integration time is given by the total amount of time the various antenna baselines spend within that pixel of Fourier space. We can transform this $\{\u_{x}, \u_{y}, f\}$ cube along the frequency/redshift dimension and map to cosmological coordinates to obtain a measurement in $\{k_{x}, k_{y}, k_{z}\}$. The uncertainty per 3D cell in $k$-space is 
\beq
v_{\rm rms} (\k) = \frac{\lambda^{2}\Tsys \sqrt{B}}{A_{e}\sqrt{\tau}},
\label{visRMSk}
\eeq
where $B$ is the bandwidth of the measurement (limited at EoR frequencies by cosmic evolution to $\sim$8~MHz), and $\tau$ is the total integration time by all baselines for the cell. 

\begin{figure}
\begin{center}
\includegraphics[width=5 in]{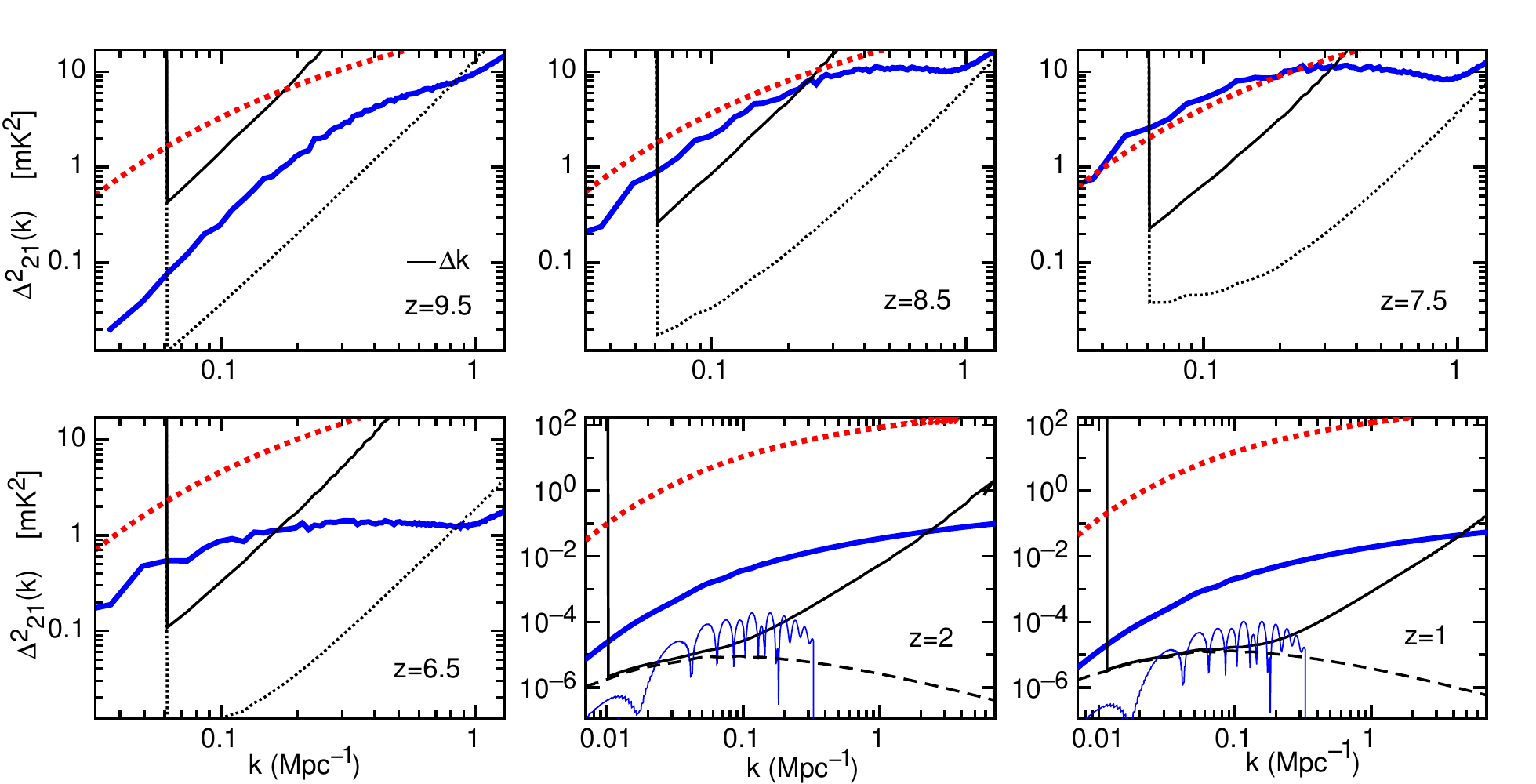}
\caption{ 
Comparison of spherically averaged 21~cm power spectrum and array sensitivity at redshifts between $z=9.5$ and $z=1$. The thick blue curves show a model for the 21~cm power spectrum signal \citep[][during reionization and the linear mass power spectrum with $x_{\rm HI}=0.01$ populating dark matter halos of $M\geq 10^{11}$M$_\odot$ for $z =$ 1 \& 2]{wyithe2009}, while the dotted red line shows the 21~cm PS assuming a fully neutral IGM as a reference. 
At $z>6$ the sensitivity is estimated over 1 field and within a 32~MHz bandpass for the MWA (dark solid lines) and for a future array with a 10-fold increase in the number of antennae tiles \citep[dark dotted lines][]{McQ+06}. At $z =$ 1 \& 2 the sensitivity is estimated over 3 fields and within a 300~MHz bandpass for CARPE (dark solid and dashed lines show overall sensitivity and the cosmic variance limit respectively.) The sensitivity is plotted within bins of width $\Delta k=k/10$ assuming 1000~hr integration in both cases. The sharp upturn at low $k$ is due to the assumption that foreground removal prevents measurement of the power spectrum at scales corresponding to a bandpass larger than $1/4$ of the total observed bandpass. To illustrate the additional sensitivity required for dark energy experiments, the baryonic oscillation component of the power spectrum is plotted at $z =$ 1 \& 2 (thin blue curves).}
\label{sensitivity}
\end{center}
\end{figure}

For a power spectrum signal, what is defined is the variance of the signal $P(k)$. In practice the measurement $I(k)$ is divided into two separate integrations of equal length, which are multiplied together to form $|I(k)|^{2}$. (Splitting the measurement eliminates the thermal noise power contribution, as the noise is independent for the two integrations but the signal is the same, M. Tegmark private communication.) The expected signal in each $k$-space cell is given by
\beq
\langle \left| I(k) \right|^{2} \rangle = \int P(k) \left| B(k) \right|^{2} d^{3}k,
\label{expSignal}
\eeq
where we have mapped the beam $B(\u,f)$ into $k$-coordinates.\footnote{Be \emph{very} careful when inserting a predicted power spectrum into Equation \ref{expSignal}, as the Fourier transform convention must be the same for all terms. There are three Fourier transform conventions in common use, each with different normalizations ($2\pi$'s), and this has led to numerous mistakes. } The thermal uncertainty per $k$-space pixel is exponentially distributed (due to squaring the measurements) and has an rms of
\beq
\left[C^{N}\right]_{\rm rms} = 2\left(\frac{\lambda^{2}\Tsys}{A_{e}}\right)^{2}\frac{B}{\tau}\delta_{ij}.
\label{thermalVar}
\eeq
In addition there is a sample variance contribution to the uncertainty
\beq
\left[C^{SV}\right]_{\rm rms} = P(k)\frac{\lambda^{2}B}{A_{e}}\delta_{ij}.
\label{sampleVar}
\eeq
For intensity mapping measurements the expected signal per $k$-cell (Equation \ref{expSignal}) and the uncertainties (Equation \ref{thermalVar} and \ref{sampleVar}) can be used in a Fisher matrix analysis to directly determine the uncertainty on the cosmological parameters (such as those for the dark energy equation of state). For EoR analyses, one typically assumes isotropy and forms a variance-weighted average of all the $k$-cells within a spherical annulus of $|k|  \pm \Delta k/2$. The uncertainty on the resulting one dimensional power spectrum bin is the weighted average of the thermal and sample variance terms (Equations \ref{thermalVar} and \ref{sampleVar}), and is proportional to $1/\sqrt{N_{k}}$ where $N_{k}$ is the number of $k$-cells averaged over. For one dimensional power spectrum, it is convenient to plot the measured values in terms of the input power spectrum by dividing both the measured signal and the uncertainty by $\int \left| B(k) \right|^{2} d^{3}k$ (from Equation \ref{expSignal}) as in Figure \ref{sensitivity}, where examples are shown both for the MWA at $z>6$ and for the CARPE at $z=1-2$.


\begin{figure}
\begin{center}
\includegraphics[width=5 in]{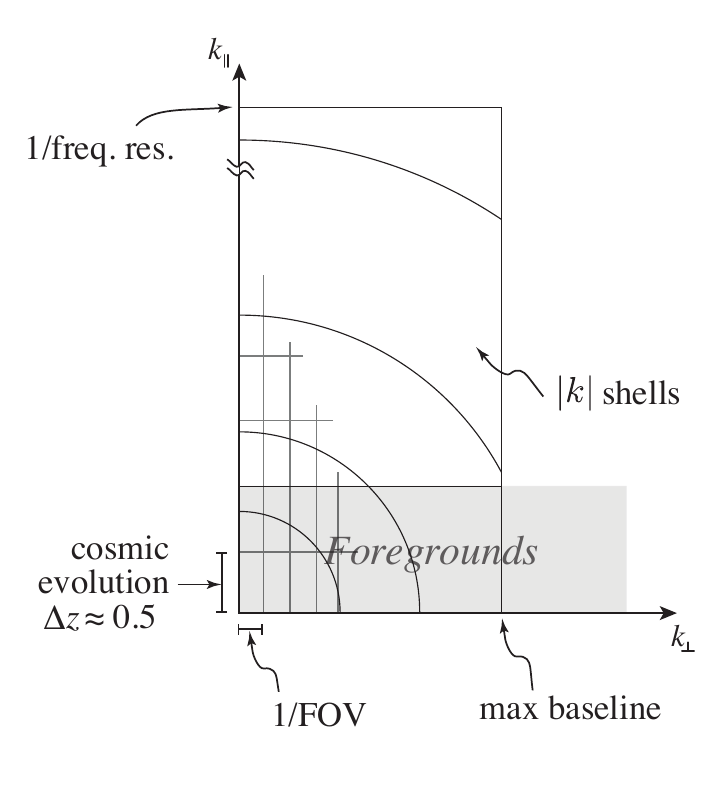}
\caption{Illustration of the $k$-space measurement space of an HI interferometer. The instrument measures the intensity in $k$-space cells, with the power spectrum intensity and uncertainty per cell described by Equations \ref{expSignal}, \ref{thermalVar}, \& \ref{sampleVar}. For EoR measurements the power spectrum intensity is averaged within $|k|$ annuli (curved regions). The maximum $\kperp$ is set by the longest antenna baseline, and the perpendicular width of the cells is 1/FOV. The maximum $\kpar$ is the inverse of the frequency resolution, and is typically much larger than the maximum angular wavenumber. The line-of-sight length of the cells is set by the inverse of the bandwidth for intensity mapping instruments, and for EoR observations bandwidth is further limited by the $\Delta z \approx 0.5$ cosmic evolution limit. Lastly, foreground contamination removes the first few $\kpar$ modes (\S \ref{foregrounds}).}
\label{kbins}
\end{center}
\end{figure}

It is instructive to look at how these relationships affect array design. Each antenna pair, over a short observation time, measures a particular angular Fourier mode at all the observed frequency channels. When we move from the $\{\u_{x}, \u_{y}, f\}$ space of the original measurements to the three dimensional $k$-space of the cosmological power spectrum $\{k_{x}, k_{y}, k_{z}\}$, each antenna baseline measures one angular $\kperp$ mode but all of the line-of-sight $\kpar$ modes. A single baseline can be envisioned as measuring all of the $k$-space cells along pencil beam parallel to the line-of-sight, as shown in Figure \ref{kbins}. This means that short baselines (small $\kperp$) still measure very small spatial scales (large $k$) along the line-of-sight $\kpar$ direction. Short baselines measure all $k$ modes, whereas long baselines only measure $|k|$ modes larger than that baseline.


Given the choice of spending more time integrating on one mode and dividing the observing time across multiple modes, you always win by integrating on one mode until you reach a signal-to-noise ratio of $\sim$1 \citep{Halverson:2002}. The thermal noise on the power spectrum decreases linearly with the integration time (Eq.\ \ref{thermalVar}), until the thermal noise is less than the signal at which point the noise is dominated by the sample variance (Eq.\ \ref{sampleVar}). In the low signal-to-noise regime, measuring two modes provides a $\sqrt{2}$ improvement, whereas measuring one mode twice as long provides a factor of 2 improvement. DASI was designed to cancel the sky rotation so antenna pairs always saw the same angular mode, increasing the integration time $\tau$ on that mode at the expense of measuring fewer modes. For HI interferometers it is impractical to cancel the sky rotation, but it is common to have many redundant antenna pairs (same separation and orientation) simultaneously measuring the same mode, thus $\tau$ for a particular mode can be greater than the total observing time. 

Even for a particular $|k|$ of interest, the experimenter can choose whether to focus the measurement in portions of the annulus more along the angular direction ($\kperp$) or more along the line-of-sight direction ($\kpar$). For EoR machines the primary signal is from a largely isotropic $P(k)$, and very short baselines provide a much higher signal-to-noise:  short baselines measure all $k$, concentrating the array provides many more redundant baselines and much higher $\tau$ on short baseline cells, and the foreground contamination is least in the $\kpar$ direction (detailed in \S \ref{astroForegrounds}). For the first generation EoR arrays nearly all the cosmological sensitivity comes from baselines shorter than a few hundred meters, and longer baselines are primarily for instrumental calibration and foreground determination.

\begin{figure}
\begin{center}
\subfigure{
\includegraphics[width = .7\textwidth]{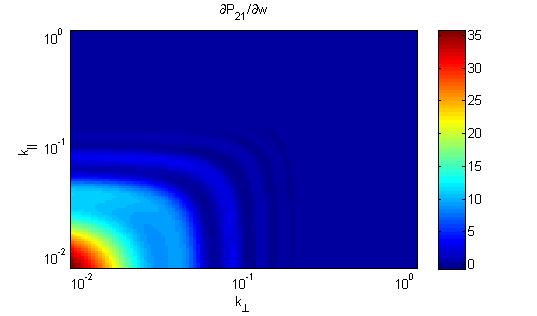}}
\subfigure{
\includegraphics[width = .7\textwidth]{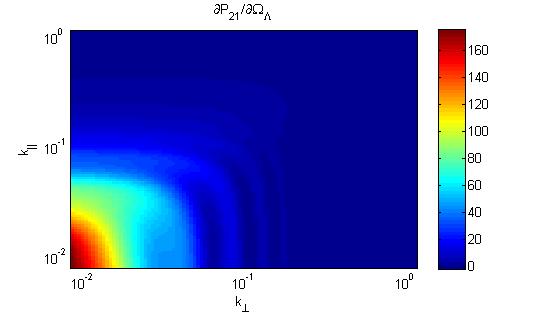}}
\caption{These plots show examples of $k$-space Fisher plots for HI observations at a redshift of 1.9 (Eli Visbal, personal communication). Each panel shows the change in the power spectrum field due to a change in one parameter---$\partial P_{\rm 21}/\partial w$ in the upper panel and $\partial P_{\rm 21}/\partial\Omega_{\Lambda}$ in the lower in mK$^{2}$. If changes in these parameters were fully degenerate the shapes of the associated changes in the power spectrum fields would be the same. In this case the $\partial w$ plot is much more symmetric (the BAO ridges appear rectangular due to the log-log axes, see Figure \ref{JuddForegrounds}) than the $\partial \Omega_{\Lambda}$ plot where there are additional features in the line-of-sight direction. These plots show where an experimenter should concentrate their $k$-space measurements to maximize the sensitivity to parameters of interest and minimize degeneracies.}
\label{EliFig}
\end{center}
\end{figure}

For 21~cm intensity mapping experiments the question is a bit more complex. Because the theory of structure formation is much more mature there is a common set of cosmological parameters ($\Omega_{\Lambda}$, $\Omega_{\nu}$, $n_{s}$, etc.). These parameters map into both the symmetric and asymmetric terms of the power spectrum in different ways (Equation \ref{DLAPS1}). The standard solution is to perform a Fisher matrix analysis for specific proposed instruments \citep[e.g.\ ][]{visbal2008} to show the parameter constraints and degeneracies of the proposed observations. However, an HI instrument designer has some freedom over both the range of scales measured and their angular distribution---by varying the antenna layout the signal-to-noise distribution in the three dimensional $k$-space can be tuned (Figure \ref{kbins}). $k$-space Fisher plots, as shown in Figure \ref{EliFig}, can help the designer of an intensity mapping instrument maximize the sensitivity to the parameters of interest, while helping break parameter degeneracies.

So far we have discussed power spectrum sensitivity exclusively, but there are a number of science applications such as Stromgren sphere imaging, non-Gaussian EoR statistics, topology/non-stationary EoR statistics, and cross-correlation with smaller area optical surveys which benefit from imaging---a signal-to-noise $>$1 per mode (corresponds to phase uncertainty on the mode less than $\sim$1~radian). This is in direct tension with the power spectrum measurements---in the power spectrum you maximize sensitivity by integrating to S/N of $\sim1$, then move to a new field (or redshift range) without ever achieving the sensitivity to create quality HI images. The boundary tends to be a bit fuzzy for HI interferometers as some $k$-cells will hit the imaging limit before others, but the tension is real. 

The power spectrum sensitivity scaling also affects the decision to track the sky vs. allowing it to drift overhead. In the low signal-to-noise regime, a tracking instrument is more efficient as it integrates longer on the same modes instead of splitting the observing time over independent modes in separate fields ($1/\sqrt{N_{k}}$). However, non-tracking antennas are simpler and cheaper to construct allowing a larger collecting area at the same instrumental cost. Both tracking and drifting approaches are being pursued by first generation HI interferometers, with GMRT, LOFAR, MWA, and CARPE tracking the sky and PAPER, CHIME and CRT drift scanning.

An example of the power spectrum sensitivity of first generation EoR and intensity mapping experiments is shown in Figure \ref{sensitivity}. The instrumental sensitivity has been calculated by determining the signal-to-noise ratio in each 3D $k$-cell, then performing a weighted average within the spherical $k$-shells (Figure \ref{kbins}) to produce the one dimensional $|k|$ power spectrum. For historical reasons there are more EoR sensitivity calculations for the MWA than other instruments, but all of the first generation machines are very similar with the limits usually just scaling by a factor of a couple. For both EoR and post-reionization instruments the sensitivity increases (lower limits) towards smaller $|k|$ until the foregrounds dominate (\S \ref{foregrounds}). During the epoch of reionization the signal evolves rapidly with decreasing redshift (compare to the red dotted fully neutral PS, and Figure \ref{Barkana2008Fig4}), while the instrumental sensitivity slowly increases as the sky temperature falls. During the peak of reionization the first generation EoR instruments can hope to see about a decade of power spectrum between the foreground limit at low $|k|$ and the sensitivity limit at high $|k|$. The increasing sky temperature makes detections of the 21~cm power spectrum beyond a redshift of $\sim$10 very challenging for first generation machines even if the evolving signal is near maximum. The majority of the information about reionization from the first generation EoR observations will come from the evolution of the power spectrum amplitude and slope \citep[][and Figure \ref{lidzfig2}]{lidz2008}. For the post-reionization instruments, much of the science case is driven by detecting the BAO signature which is plotted in the last two panels. While the BAO signal is much weaker than the reionization signal, the intensity mapping machines are helped by the much lower sky temperature at these redshifts.

The first generation EoR and intensity mapping experiments will also be able to image the largest scale structures. In Figure \ref{sensitivity} the sample variance contribution to the power spectrum is plotted as a dashed line, and starts to limit the sensitivity at the largest scales. While from a power spectrum perspective this means one should start observing additional fields, it means one can start to image at the scales where the sample variance is limiting. Figure \ref{Wyithe1} shows the imaging capability of a large quasar Stromgren sphere for a first generation EoR machine. It should be noted that imaging sensitivity should be calculated on a per-mode basis just as the power spectrum is---the radiometer equation assumes a uniform antenna distribution and is dramatically wrong for the very centrally condensed antenna distributions of most 21~cm fluctuation machines \citep{GW08}. 


\subsection{Interferometric calibration}
\label{calibration}
All of the HI experiments feature compact antenna configurations and high survey speeds to maximize the sensitivity relations detailed in the previous section. As we will detail in \S \ref{foregrounds}, successful separation of the bright astronomical foregrounds from the very faint EoR and BAO signals will require extraordinarily precise instrumental calibration. EoR instruments are pushing the state of the art in the calibration of radio instruments \citep{Mitchell:2008p3967,Pen:2008p3969,Nijboer:2007p4011,Parsons:2009p3997}, and it is in calibration that the craft of instrument building comes to the fore.


Traditional calibration determines a single per-antenna gain and phase using observations of a bright isolated point source. This is usually performed using the selfcal algorithm \citep[detailed in][]{SynthImagII} which assumes all of the flux comes from the dominant source and adjusts the complex antenna gains to maximize this flux, solving for a single complex gain per antenna (amplitude and time delay). This single complex gain calibration works well in the limit of all antennas being identical and the atmospheric propagation delay being constant across the field-of-view. For an instrument like the VLA (Very Large Array) the antennas are regularly adjusted to keep their figures close to identical and the narrow field of view naturally limits angular variation in the atmospheric propagation delays.

While gain and time delays generated after the antenna couples the incident radiation into the RF cable (cable losses, temperature dependent amplifier gain and filter characteristics, mixer phase delays, etc.)\ do affect all of the incoming electric signal equally, effects which occur within the optics of the antenna can impart direction-dependent amplitude and delay changes. The direction-dependent gain can be envisioned as a envelope which attenuates the true incident electric field, or more physically it can be represented in the Fourier plane as the efficiency (and delay) of radiation hitting each segment of the antenna of coupling into the RF cable. The Fourier representation is more commonly known as the holographic antenna map \citep{Scott:1977p2425}, and is typically produced by using a reference antenna to look at a very bright astrophysical source while the antenna under calibration is raster scanned so the apparent amplitude and phase of the source can be measured in all look directions. This gain map is Fourier transformed to produce an image of the antenna surface.
Holographic antenna maps reveal important physical constraints on the antenna calibration problem:  only electric field hitting the surface of the antenna can couple in so the gain cannot vary with direction faster than $\lambda/D$ (wavelength and antenna diameter);\footnote{reflections from outside the antenna, curved sky effects, and Fresnel diffraction of the antenna's three dimensional structure can produce small contributions from outside the antenna diameter \citep{Morales:2009p3737}.} and features smaller than $\sim\!\lambda/2$ are averaged over by the spatial coherence of the electric field ($\theta \lesssim \pi/2$, a 1 m wave does not see a 1 cm bolt). This provides an upper limit  of $\sim\!4A/\lambda^{2}$ on the number of parameters needed to describe the single frequency calibration of an antenna---one complex number for each $\lambda^{2}/4$ patch of the antenna surface. In practice the number of calibration parameters may be significantly less if there are typical modes of deformation (e.g. gravity sag of a dish, dipole based gain factors), but these are best conceptualized as additional constraints on the characteristics of the antenna reception surface.


The scales of the atmospheric/ionospheric phase delays depend on the amplitude and inner and outer scales of the turbulent Kolmogorov cascade, plus additional modes such as large scale Traveling Ionospheric Disturbances (TIDs). There are three different regimes of atmospheric distortions which concern HI interferometry \citep[following][starting with his regime 2]{Lonsdale:2005p4452}.  In regime 2, the antenna baselines are long compared to the ionospheric scales, but the field-of-view is small so the length scale at ionospheric heights is small compared to the minimum distortions. In this regime sources scintillate, but the scintillation pattern is the same for all sources in the field and it can be calibrated with a single ionospheric time delay per antenna. Regime 2 is commonly encountered with VLA observations and is well corrected by the traditional selfcal algorithm. In regime 3 the antenna baselines are shorter than the ionospheric distortion scales, but the field of view is large so the apparent source positions are refracted in different amounts across the field but do not scintillate. This can be corrected with a `rubber sheet' correction to map sources from their observed location back to their true location, and is the typical regime for EoR observations. Regime 4 is the most challenging situation, where both the antenna baselines and field of view are large compare to ionospheric distortions, leading to a direction-dependent scintillation of sources. The correction of these regime 4 distortions will be needed for high resolution imaging by LOFAR and the LWA (Long Wavelength Array) and are at the forefront of ionospheric calibration research \citep{vanderTol:2007p4454}.

Mathematically correcting interferometric radio data is identical to optical/IR adaptive optics, with regimes 2 and 4 corresponding to narrow and widefield adaptive optics respectively. The Fried length for ionospheric distortions (length over which phase delay is $\ll 2\pi$, i.e.\ refraction but no scintillation) is typically several kilometers with timescales of minutes, but can vary by orders of magnitude depending on the ionospheric conditions and the frequency of observation. Of additional concern to HI cosmology observations is non-diagonal noise contributions from ionospheric distortions. While the noise of individual mode measurements (visibilities) is usually independent, the stretching and compressing of ionospheric distortions can create the spatial analog of intermodulation distortion, mixing spatial modes before they are measured by the array \citep{Morales:2009p3737}. This introduces non-diagonal correlations into the visibility correlation matrices, and it is unclear what effect this may have on extracting cosmological signals.

The direction-dependent antenna gain and the ionospheric distortions can be corrected if enough calibrators can be observed to determine the calibration parameters. Because the radiation wavelength is on the meter scale, it is difficult to construct an anechoic chamber large enough to get into the antenna far field, and no one has succeeded in designing a precision balloon, satellite, or tower based calibration system (any conductive support or wire is reflective, and it must raster through the antenna field-of-view). For these reasons all of the first generation EoR instruments are relying on compact astrophysical objects as their calibration sources. The difficulty with astrophysical calibrators is source isolation---when the emission from one source is selected it contains some emission from sources across the field due to the finite point spread function (Figure \ref{EIsquares} panel $c$). The cross-leakage of calibrator sources adds systematic errors that limit the precision to which the calibration parameters can be solved for.

The four EoR observatories have taken different approaches to calibration, and are all pushing the state-of-the-art in radio calibration. Briefly the approaches are:
\begin{itemize}
  \item GMRT is using a novel pulsar based calibration approach \citep{Pen:2008p3969}. Using a customized correlator they separate visibilities from when the pulsar is on from when it is off. By subtracting the on and off data they are left with only the pulsar emission, beautifully isolating the emission of their calibrator from all other sources. Unfortunately the space density of bright pulsars means there is only one calibrator in their field of view. This is less of a problem for the GMRT as it has the smallest field of view of the EoR machines, but it restricts the ability to perform direction-dependent calibrations.
  \item PAPER calibration relies on a very stable antenna gain pattern \citep{Parsons:2009p3997}. The PAPER antenna is designed to have a predictable gain response, and as it only looks straight up the response is very stable. Effectively this reduces the number of calibration parameters that need to be solved for.
  \item MWA calibration simultaneously fits the gain for $\sim$100 calibrator sources across the antenna field of view and sidelobes, effectively producing holographic antenna maps and an ionospheric refraction map every 8 seconds \citep{Mitchell:2008p3967}. The 512 antennas of the MWA provide an excellent point spread function, improving the source isolation between calibrators at the cost of an enormous data rate. The data rate necessitates that the calibration be performed in real time, as the data volume is too large to archive for long observations.
  \item LOFAR has fewer `virtual' antennas (stations) than the MWA, and can thus afford to save all of the visibility data for offline processing. However, this reduces the source isolation. LOFAR calibration uses a full maximum likelihood solution of the calibration parameters coupled with their EoR foreground subtraction and a three-dimensional ionospheric model to work around the calibration leakage issue \citep{Labropoulos:2009p4015,Nijboer:2007p4011}.
\end{itemize}

So far the 
intensity mapping instrument designs have concentrated on achieving sufficient signal-to-noise, but we expect calibration to become a focus as the designs mature. In this review we have concentrated on measuring the calibration parameters, but how to use the calibration in the data analysis pipeline should also be considered. The interested reader is referred to the overview by \cite{Rau:2009p4084} and literature on the analysis of direction-dependent calibrations through either image faceting \citep{Schwab:1984p4135,SynthImagII} or more recent software holography techniques  \citep{Bhatnagar:2008p3407,Morales:2009p3737}.

\subsection{Foreground removal}
\label{foregrounds}

The primary challenge facing HI measurements is foreground subtraction. Astrophysical sources such as Galactic synchrotron radiation and extra-galactic radio galaxies provide a sea of foreground emission $\sim$5 orders-of-magnitude brighter than the EoR and 
post reionization redshifted 21~cm fluctuation signals. In addition there are strong terrestrial and satellite broadcasts and a number of instrumental effects. These unwanted contaminants must be removed to high precision to detect the cosmological HI signals.

While almost all the literature concentrates on the EoR, it can be translated directly to higher frequency BAO observations. Most of the foreground work remains theoretical, and we expect the techniques will change dramatically over the next couple of years as they are confronted with real data. We are just starting to see the first deep integrations at EoR frequencies by \cite{Ali:2008p4242}, \cite{Pen:2008p3969} and \cite{Schnitzeler:2009p4246}, and these early observations are already having a strong impact.

The early foreground subtraction literature concentrated on astrophysical foregrounds that might produce a fundamental barrier to HI observations. In the past few years research has shifted to observational effects such as chromatic point spread functions and imperfect polarization purity, and how they complicate the removal of the astrophysical foregrounds. We will start with a brief review of astrophysical foregrounds in \S\ref{astroForegrounds}, then will concentrate on the more recent observational foreground issues in \S\ref{obsForegrounds}.

\subsubsection{Astrophysical foregrounds}
\label{astroForegrounds}


The major astrophysical foregrounds are synchrotron and free-free emission from our Galaxy and a confused sea of extra-galactic radio sources. The angular fluctuations in the Galactic emission at high latitude are quite small \citep{deOliveiraCosta:2008p3515} and largely resolve out in interferometric observations.\footnote{There are no zero-spacing measurements for technical reasons, so the $k$-space $\delta$-function at zero (mean intensity) is not measured. However, the unresolved Galactic emission can still dominate the system temperature \S\ref{InterferometryIntro}.} However, at arcminute scales extra-galactic radio galaxies produce a confused and highly structured surface of emission that is only a few times fainter than the Galactic emission.  

The Galactic synchrotron emission has a typical spectral index of $f^{-2.6}$, and \cite{deOliveiraCosta:2008p3515} has compiled low frequency observations into a comprehensive model of the large scale galactic intensity emission (Figure \ref{AngSynch}). While the Galactic intensity is spatially very smooth, it is also polarized and Faraday rotation introduces significant structure in polarized maps at arcminute scales. Observation by \cite{deBruyn:2006p4286} and \cite{Schnitzeler:2009p4246}  with the Westerbork Synthesis Radio Telescope (WSRT) at 315--388~MHz (redshift 5.5--4.6) have shown $\sim$3~K polarized structure on arcminute scales and $\sim$7~K fluctuations at the 150~MHz band targeted by EoR observations \citep{Bernardi:2009p4250}. However, most of the published WSRT observations have focused on the Fan region which is known to have significant polarization structure. Observations by \cite{Pen:2008p4020} with the GMRT in a different 150~MHz window show significantly less polarized emission. From current observations it is unclear how much polarized structure there will be in the windows of interest for EoR measurements, though much larger surveys of polarized foreground emission will be released soon (Gianni et al.\ personal communication). 

The extragalactic foreground emission ranges from extremely bright radio galaxies (e.g.\ Cas A) down to a confused sea of faint sources. On the faint end there are hundreds of sources per angular pixel---\emph{all} pixels contain foreground emission. Our expectations for the extragalactic emission come primarily through increasingly detailed simulations by \citet{2004MNRAS.355.1053D,Bowman:2009p4044}, and in particular the comprehensive models by \citet{Jelic:2008p2041}. The confusion level foreground has additional spectral features due to the co-addition of many sources along the same line-of-sight with different spectral indices and curvature points, though the luminosity distribution is flat enough that the emission tends to be dominated by the brightest source in each angular pixel.

\cite{2002ApJ...564..576D} and \cite{Oh:2003p2809} showed that the clustering and fluctuations of radio galaxies presents an insurmountable obstacle to angular HI power spectrum measurements. The key to foreground subtraction is the spectral smoothness of the synchrotron and free-free emission \citep{2004MNRAS.355.1053D,mcquinn2007,Gnedin:2004p2873,Santos:2005p2174,Zaldarriaga:2004p227,Morales:2004p803}. The fluctuations in the HI density and spin temperature along the line-of-sight direction (redshift) create small spectral fluctuations which are superimposed on the smooth synchrotron and free-free foreground spectrum.  First generation HI instruments can be thought of as looking for small spectral fluctuations due to changes in the HI density along the line-of-sight. While the foreground emission is not perfectly flat due to synchrotron self-absoption, the admixture of sources with different spectral indices, and other effects, all of the foreground spectral features are very smooth. A number of techniques have been developed for fitting out the slowly varying foreground emission along each line of sight, including polynomials \citep{mcquinn2007,Jelic:2008p2041}, double log polynomials \citep{Wang:2006p503}, and non-parametric techniques \citep{Harker:2009p4243}. All of these methods work well, and qualitatively, fit out the first few line-of-sight Fourier modes where the foreground dominates, leaving the more rapid line-of-sight HI fluctuations.

It is instructive to look at the foregrounds in the three dimensional $k$-space of Figure \ref{kbins}. The Galactic emission is very smooth both spatially and spectrally, and only adds power at very small $|k|$. The extra-galactic emission has a lot of angular $\kperp$ power due to fluctuations in the number of galaxies per angular pixel and galaxy clustering, but their spectra are also very smooth and fall rapidly after a couple of modes in $\kpar$. Qualitatively the foreground removal techniques remove the first $\sim$3 $\kpar$ bins along the line of sight, leaving the higher $\kpar$ bins largely free of the foreground contamination. 

In addition, the functional forms of the foregrounds and expected signals differ:  the cosmological signal is approximately spherical in shape due to the combination of spatial isotropy and a redshifted emission line (\S \ref{HIsensitivity}), whereas the foregrounds have a separable-axial symmetry ($\sim$cylindrical) because there is no relationship between the angular and line-of-sight/frequency emission \citep{Morales:2004p803,Zaldarriaga:2004p227}.

The difference in the three dimensional $k$-space symmetries provides an important handle for distinguishing foregrounds from the HI signal. The only foregrounds which can mimic the symmetry of the HI signal are radio lines at high redshift. There was concern that redshifted radio recombination lines could provide just such a confusing foreground, but more recent analysis predicts this will be a very small contribution \citep{Furlanetto:2006p341}. Symmetry-based subtraction algorithms can further reduce foreground contamination and determine the systematic errors due to foreground subtraction residuals \citep{Morales:2006p147}.


It is possible to partially recover the angular modes (small $\kpar$ component) on large scales where the measurement pushes into the $SN>1$ per mode imaging regime. Simulations by \citet{GWPO08} and \citet{Furlanetto:2006p341} show that large ionized Stromgren spheres around quasars can be imaged in the presence of strong foregrounds, though foreground removal does decrease the contrast of the image (Figure~\ref{Wyithe1}). 

The general consensus is that there are no known astrophysical foregrounds which present a fundamental impediment to HI power spectrum or imaging observations. However, no instrument is perfect and removing astrophysical foregrounds from real world observations remains the primary challenge of both EoR and intensity mapping measurements. In the literature the discussion of foreground removal has shifted from the fundamental astrophysical foregrounds discussed in this section, to the removal of foregrounds from upcoming observations which we discuss next.

\subsubsection{Observational foregrounds}	
\label{obsForegrounds}

While astrophysical foregrounds can be removed in principle using the smooth spectral properties, small observational contaminants can mix bright astrophysical and terrestrial foregrounds into the spectral domain and mask the faint HI spectral fluctuations. Observational contaminants include ionospheric refraction and scintillation, non-linearity in the analog RF system, chromatic PSF and FOV, and calibration errors in the direction-dependent gain and polarization response of the antennas. These errors can combine with astrophysical and terrestrial foregrounds to create second-order contaminants---an effect commonly called mode mixing.

\begin{figure}
\begin{center}
\includegraphics[width=5.25in]{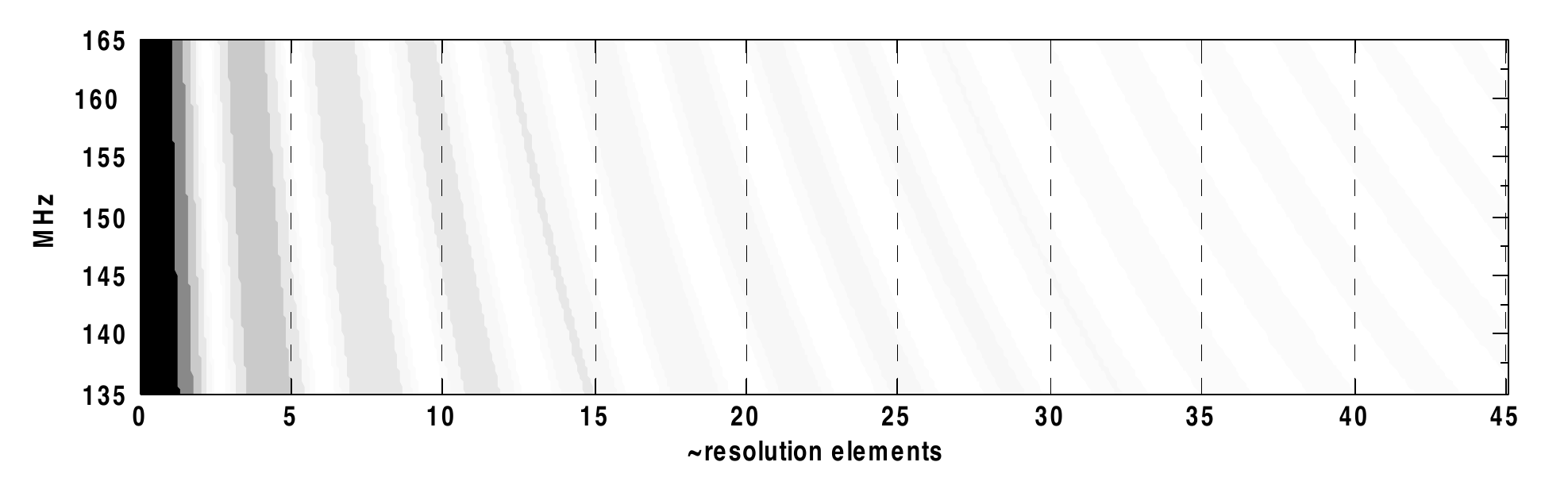}
\caption{This cartoon illustrates mode-mixing between a frequency dependent point spread function (PSF) and a mis-subtracted point source. At the origin of the horizontal axis a source has been mis-subtracted, and the residual flux ripples across the image due to the array's point spread function. Since the PSF scales with frequency (center frequency times fractional bandwidth, lowest frequency at bottom and highest at top) the positions of the radial intensity peaks change with frequency resulting in diagonal bands of spurious signal. At lines of sight away from the mis-subtracted source this leads to spectral fluctuations in the line-of-sight, as indicated by the contribution along the vertical dashed lines. The frequency dependent PSF is inherent in the measurement, and mixes a pure spatial component (mis-subtracted sources or mis-calibrated antenna response) into the frequency direction. In addition, as the frequency slope of the contamination depends on the distance to the residual sources, a wide range of line-of-sight $k$-modes are contaminated.}
\label{modemixFig}
\end{center}
\end{figure}

As an example of mode mixing, let's consider a mis-subtracted point source (e.g. a radio galaxy) interacting with the chromatic instrumental PSF as shown in Figure \ref{modemixFig}. The PSF of all HI interferometers is inherently chromatic---the apparent separation of the antennas in wavelengths changes as a function of frequency (Figure \ref{EIsquares} panels $c$ \& $d$). If a source is mis-subtracted, either because the flux is not known precisely, the instrumental gain of all the antennas towards the source is not known accurately (error in knowledge of the angular PSF), or the frequency dependence of the antenna gains is not known accurately (error in frequency dependent changes of the PSF), there will be residuals in the image cube proportional to the product of the errors in the source flux and the PSF or the product of the source intensity and the errors in the instrumental calibration. Because the angular position of these contaminants changes as a function of frequency, they introduce spectral fluctuations along lines-of-sight away from the source location. Effectively this has mixed a purely spatial foreground---a smooth spectrum point source---into the frequency domain, complicating the foreground subtraction techniques discussed in the previous section. 

The experimental emphasis on instrumental and atmospheric calibration (\S \ref{calibration}) is an effort to mitigate these mode mixing contributions to the foreground subtraction problem. Observational foregrounds are characteristically a product between an instrumental term and an astrophysical or terrestrial foreground. The product of our calibration uncertainties and our foreground uncertainties must be less than the expected EoR signal, or better than a part in $10^{5}$ for most mode mixing terms. While mitigation of observational artifacts has been a significant design driver for EoR and intensity mapping observatories, much of this work is only now working its way into the literature (they're busy building instruments). In this section we will briefly review the main sources of observational foreground contamination and the techniques being used to reduce their effect.

\emph{Ionospheric calibration.}  Phase delays in the ionosphere lead to refractive displacements (and at times scintillations) in the apparent location of bright sources. Errors in the ionospheric calibration can lead to the source subtraction being slightly offset from the apparent source location. This leaves residual flux in the image cube with a characteristic chromatic PSF pattern, mixing the angular error in the source location into the frequency domain. As ionospheric displacement scales as $\lambda^{2}$, this is a larger problem for EoR observatories than post-reionization arrays. All of the EoR arrays are located at mid geomagnetic latitudes (between the magnetic pole and equator) where the ionosphere is better behaved. The low PSF sidelobes of the MWA reduce both the contamination from mis-subtractions and the systematic uncertainties in the apparent source locations \citep{Mitchell:2008p3967}, while LOFAR uses its long baselines to create a full 3D model of the ionosphere \citep{Intema:2009p4453}. 

It has been suggested that as the ionospheric fluctuations are statistical, they could be ignored and this effect would average out (effectively the foreground sources would all be smeared in a similar way, Zaldarriaga \& Tegmark personal communication). However, we have very little data on the statistical distribution of ionospheric distortions and it is unclear if they will average out sufficiently the over course of proposed EoR observations.

\emph{RFI.} Radio frequency interference from terrestrial and satellite transmitters can be extraordinarily bright---a one Jansky astronomical source is only $10^{-26}$~W/m$^{2}$~Hz. Most transmissions are characteristically narrow in frequency and can be excised in the frequency domain.\footnote{Digital transmissions do not have a dominant carrier wave and transmit much more information, thus have a broader frequency signature and fill much more of their frequency allocation.} However, non-linearities in the analog receiving system introduce frequency mixing (intermodulation distortion), producing signals at the sum and difference of the RFI frequencies. Recursive mixing of the RFI and the mixing products can create a forrest of faint frequency contaminants. Several projects have selected remote areas of Western Australia (PAPER, MWA, CARPE) or the Sahara (CRT) to reduce the amplitude of the RFI, while LOFAR and the GMRT have concentrated on advanced RFI subtraction algorithms \citep{Pen:2008p3969}, very linear front ends, and higher ADC bit depths.


\emph{Astronomical point sources.}  The flux of smooth spectrum astrophysical sources can mix into the frequency domain as described earlier (Figure \ref{modemixFig}). For bright astronomical sources the source flux and spectrum can be well determined, and most of the mixing comes from errors in the instrumental calibration of the antennas. In particular, the direction-dependent gain calibration of each antenna must be well characterized so the PSF towards each bright source is well known. 

\begin{figure}
\begin{center}
\includegraphics[width=\textwidth]{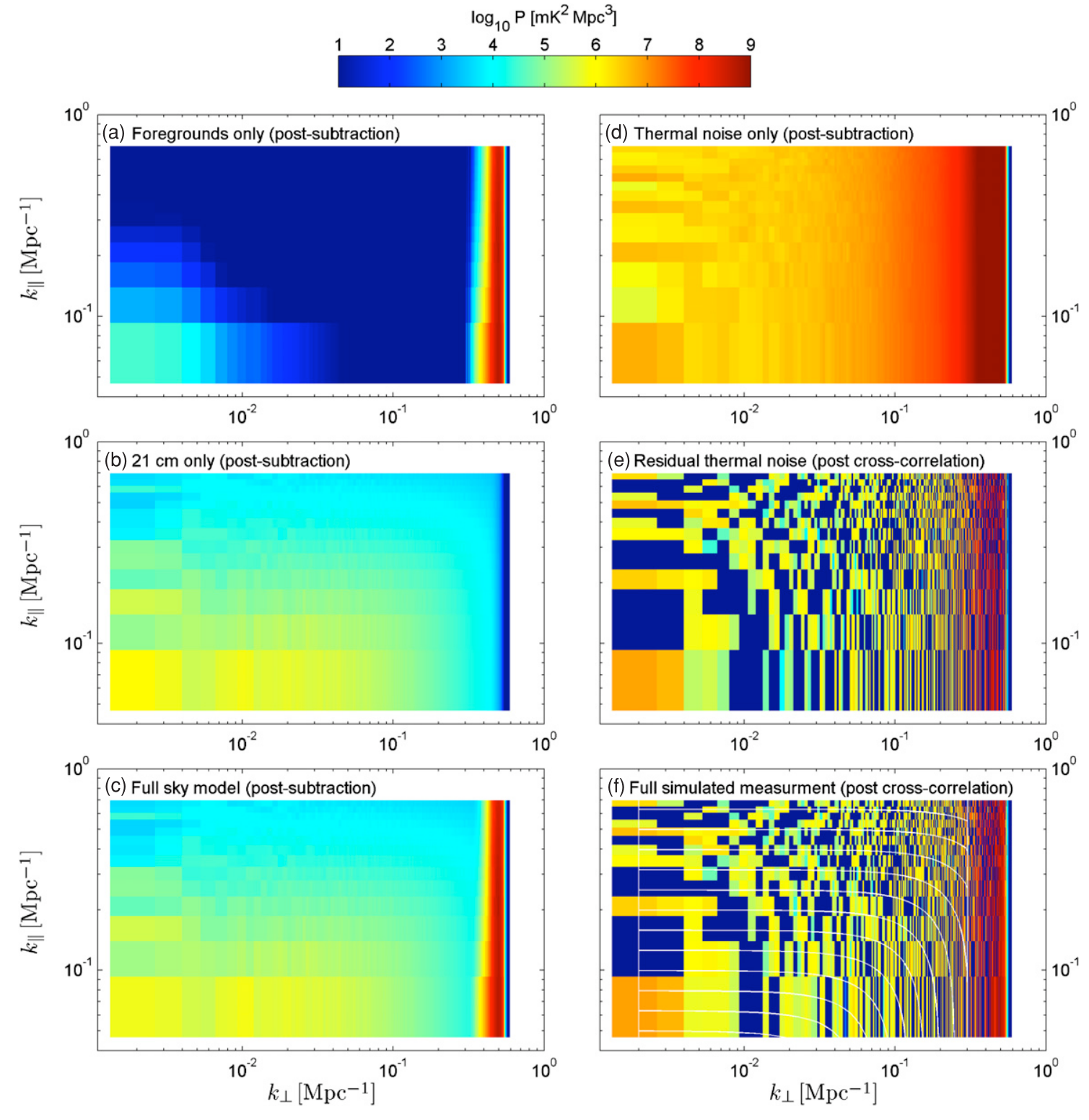}
\caption{Two-dimensional power spectrum of postsubtraction residuals for confusion level sources including chromatic PSF mode-mixing \citep[from ][]{Bowman:2009p4044}. For the left column of panels, the thermal noise has been artificially removed. It is evident that the foreground-only residuals following subtraction are much lower than the 21 cm signal in the range $\kperp \lesssim 3 \times 10^{-1}$, beyond which corresponds to when the $uv$ coverage of the MWA becomes sparse. Panel (c) confirms that the 21 cm signal dominates the recovered power from the full-sky model since it appears nearly identical to panel (b). The white arcs in panel (f) illustrate the spherical shells of constant $|k|$ that are used to determine the spherically averaged one-dimensional power spectrum. Note it is only after averaging in $k$-shells that the PS detection becomes possible.}
\label{JuddForegrounds}
\end{center}
\end{figure}

For the confusion level sources it is much harder to determine the true source fluxes, as the observed brightness in an angular pixel is a combination of the true source intensity and the sum all other sources times the array PSF. In this case it is the uncertainty in the source fluxes which dominates the mode mixing. \cite{Bowman:2009p4044} and \cite{Liu:2008p3892} have looked at this contamination for image and $uv$-plane foreground subtraction approaches respectively. Particular care must be taken to ensure that the algorithm fitting the smooth frequency emission of the confusion level sources is not biased by the chromatic PSF contributions. Figure \ref{JuddForegrounds} shows the residual $k$-space contributions from the image space foreground fitting. While these algorithms are specific to the MWA and its excellent baseline coverage, these studies suggest that the chromatic PSF and confusion level source mode-mixing can be removed to sufficient accuracy.

\emph{Galactic polarization.} While the EoR and post-reionization cosmology signals are unpolarized, the polarized Galactic emission can leak into the unpolarized intensity measurement through instrumental polarization calibration errors. Faraday rotation by the interstellar magnetic field creates a $f^{-2}$ rotation of the observed polarized Galactic synchrotron emission \citep[$\phi \propto \lambda^{2}$, ][]{deBruyn:2006p4286,Schnitzeler:2009p4246,Bernardi:2009p4250,Pen:2008p3969}. An enhancement in the instrument response at a particular polarization angle will create structure in the frequency direction as the rotating polarization angle beats with the polarized instrument response. Instrumental calibration errors mix the Stokes I, Q, U, \&V maps, and the Galactic synchrotron emission has a lot of spatial and frequency structure in Q  \& U. The polarized Galactic emission is a few K on arcminute scales (\S \ref{astroForegrounds}), and there are often multiple rotation measures along each line-of-sight providing emission components which rotate at different rates and interfere to create additional frequency structure. 

In addition to precision polarization calibration, it is possible to use the characteristic $\lambda^{2}$ dependence to separate the polarized Galactic emission from the EoR signal. By going to rotation measure space---Fourier transform of $\lambda^{2}$---much of the polarized Galactic emission will resolve into components at specific rotation measures (Gaensler, personal communication). These contaminants can then be subtracted, as the rotation measure space is largely orthogonal to $\kpar$.

The EoR observatories are at the forefront of radio calibration, and mode-mixing is likely to determine the success of the first generation of HI measurements. If it were not for the presence of strong foreground emission, the calibration requirements of HI machines would be rather modest. However, the mixing of the foreground emission into the frequency domain via instrumental effects places stringent limits on the calibration of first generation 21~cm fluctuation observatories. 


\section{Conclusions}

The study of reionization and cosmology stands on the verge of a new era. In the next couple of years we will see the first sensitive attempts to measure 21~cm fluctuations from the EoR, and the start of construction of the first intensity mapping instruments. Progress towards understanding 21~cm fluctuations has been proceeding rapidly along parallel theoretical and observational paths.

On the theoretical side, computer simulations of reionization have progressively
achieved larger volumes and dynamic range, and have now reached a point of maturity where reliable (though not yet precise) predictions can be made. These simulations (an example of which is shown in Figure~\ref{Trac1}) describe a hydrogen reionization that started with HII regions forming around the first galaxies. These HII regions grow to surround groups of galaxies, with reionization completing once the regions overlap and occupy most of the cosmological volume, leaving only dense islands of neutral gas. 
The dominant remaining numerical issues are concentrated around the transport of radiation from the sources through the IGM. Numerical simulations use a range of approximations to address this problem, but have not yet converged to a common solution. Such a convergence is a critical milestone on the path towards simulations that will be used to quantitatively interpret future observations of 21~cm fluctuations. With this goal in mind the Cosmological Radiative Transfer Codes Comparison Project \citep{ilievc2006} has compared a number of different radiative transfer codes on a set of standardised scenarios, including a cosmological density field. Although some differences remain that can be traced to technical details of the differing approximations and algorithms, the results are encouraging with general agreement on the properties of the resulting ionization field. These converged  simulations will provide the benchmark for more efficient analytic and semi-numerical models whose flexibility allows the exploration of more extremes of parameter space. As the numerical calculation of ionization structure becomes more accurate for a given model of high redshift galaxies, attention will need to focus on the properties of the sources driving reionization. In particular, self consistent account of all contributions to the development of ionized structure, in addition to stellar ionization from galaxies, will need to be considered to draw robust conclusions from upcoming 21~cm observations. 

Observationally, there are four EoR machines planning to start sensitive 21~cm power spectrum observations in the next two years, and there are several intensity mapping instruments under active development. While the bright astrophysical foregrounds should not be a fundamental impediment to HI observations, they place very tight constraints on the observational precision needed to keep the foregrounds from mixing into the faint 21~cm fluctuations. The 21~cm fluctuation machines are pushing the state-of-the-art in low frequency calibration, digital data processing, high dynamic range imaging, and widefield observations. The impending EoR measurements will teach the observational community how to perform precision cosmological measurements at low radio frequencies. This experience will be invaluable for both subsequent EoR measurements and first generation intensity mapping machines. 

The past decade has seen a maturing of cosmology into a precision science via the development of massive galaxy redshift surveys and high resolution CMB maps. Observations of 21~cm fluctuations promise to revolutionize our understanding of early galaxy formation and provide a powerful new tool for the next generation of precision cosmology measurements.



\vskip 0.2in

\section*{Acknowledgements}
 We wish to acknowledge the very helpful reviews by Steve Furlanetto and Leon Koopmans, and thank Gianni Bernardi, Bryna Hazelton, and Marcelo Alvarez for their input and comments on early versions of this review. JSBW would like to thank Avi Loeb for sparking his interest in this topic and for his ongoing collaboration. MFM would like to thank Jacqueline Hewitt, Colin Lonsdale, and Roger Cappallo for teaching him precision interferometry, and he would like thank all the scientists who have joined him in long discussions on the fine points of 21~cm observations and instrumentation, often over pints of beer. Our work is supported by the Australian Research Council (JSBW) and National Science Foundation grant 0847753 (MFM).

\bibliographystyle{Astronomy}
\bibliography{morales,wyithe,misc}

\end{document}